\documentclass[fleqn,usenatbib]{mnras}

\usepackage{newtxtext,newtxmath}

\usepackage[T1]{fontenc}
\DeclareRobustCommand{\VAN}[3]{#2}
\let\VANthebibliography\thebibliography
\def\thebibliography{\DeclareRobustCommand{\VAN}[3]{##3}\VANthebibliography}

\usepackage{lineno}

\usepackage{graphicx}	
\usepackage{amsmath}	

\title[Redshifts from EBL and {\it Fermi}-LAT data]{Constraints on redshifts of blazars from extragalactic background light attenuation using {\it Fermi}-LAT data}

\author[Dom\'inguez et al.]{
Alberto Dom\'inguez,$^{1}$\thanks{E-mail: alberto.d@ucm.es}
Mar\'ia L\'ainez,$^{1}$\thanks{E-mail: malainez@ucm.es}
Vaidehi~S. Paliya,$^{2}$\thanks{E-mail: vaidehi.s.paliya@gmail.com}
Nuria \'Alvarez-Crespo,$^{1}$
Marco Ajello,$^{3}$\newauthor
Justin Finke,$^{4}$
Mireia Nievas-Rosillo,$^{5,6}$
Jose Luis Contreras,$^{1}$
Abhishek Desai$^{7}$
\\
$^{1}$IPARCOS and Department of EMFTEL, Universidad Complutense de Madrid, E-28040 Madrid, Spain\\
$^{2}$Inter-University Centre for Astronomy and Astrophysics (IUCAA), SPPU Campus, 411007, Pune, India\\
$^{3}$Department of Physics and Astronomy, Clemson University, Kinard Lab of Physics, Clemson, SC 29634-0978, US\\
$^{4}$U.S. Naval Research Laboratory, Code 7653, 4555 Overlook Avenue SW, Washington, DC 20375-5352, USA\\
$^{5}$Instituto de Astrof\'isica de Canarias, E-38205 La Laguna, Tenerife, Spain\\
$^{6}$Universidad de La Laguna, Dept. Astrof\'isica, E-38206 La Laguna, Tenerife, Spain\\
$^{7}$Department of Physics and Wisconsin IceCube Particle Astrophysics Center, University of Wisconsin-Madison, Madison, WI 53706, USA
}

\date{Accepted XXX. Received YYY; in original form ZZZ}

\pubyear{2022}

\begin{document}
\label{firstpage}
\pagerange{\pageref{firstpage}--\pageref{lastpage}}
\maketitle

\begin{abstract}
The extragalactic high-energy $\gamma$-ray sky is dominated by blazars, which are active galactic nuclei with their jets pointing towards us. Distance measurements are of fundamental importance yet for some of these sources are challenging because any spectral signature from the host galaxy may be outshone by the non-thermal emission from the jet. In this paper, we present a method to constrain redshifts for these sources that relies only on data from the Large Area Telescope on board the {\it Fermi Gamma-ray Space Telescope}. This method takes advantage of the signatures that the pair-production interaction between photons with energies larger than approximately 10~GeV and the extragalactic background light leaves on $\gamma$-ray spectra. We find upper limits for the distances of 303 $\gamma$-ray blazars, classified as 157 BL Lacertae objects, 145 of uncertain class, and 1 flat-spectrum-radio quasar, whose redshifts are otherwise unknown. These derivations can be useful for planning observations with imaging atmospheric Cherenkov telescopes and also for testing theories of supermassive black hole evolution. Our results are applied to estimate the detectability of these blazars with the future Cherenkov Telescope Array, finding that at least 21 of them could be studied in a reasonable exposure of 20~h.
\end{abstract}




\section{Introduction}\label{sec:intro}
Active galactic nuclei that have jets pointing in the direction of our line of sight are known as blazars. These sources dominate the extragalactic sky at the energies detected by the Large Area Telescope (LAT) on board the {\it Fermi Gamma-ray Space Telescope}, typically around 1~GeV \citep{3FHL,4FGL}.

About 50\% of the blazars that are catalogued by the LAT collaboration do not have a redshift \citep{4LAC,4FGL-DR3}. This may be caused by a lack of observations, or the absence of spectral features in their optical spectra. However, distance information is critical for different science cases such as (1) building the luminosity function of blazars and understanding their evolution \citep[e.g.,~][]{ajello12_fsrq,ajello14,ajello15,sheng22}, (2) studying emission mechanisms and physical properties \citep[e.g.,~][]{vandenberg19,paliya18,paliya21,nievasrosillo22}, or (3) selecting potential targets for imaging atmospheric telescopes including the Cherenkov Telescope Array \citep[e.g.,~][]{hassan17,paiano21}. Measuring these redshifts remains a challenge despite the important efforts being made to alleviate this problem \citep[e.g.,~][]{paggi14,massaro15,landoni15,alvarez-crespo16,klindt17,paiano17,marchesini19,desai19,paliya20,rajagopal21,goldoni21,olmo-garcia22,kasai23}.

Given the observational difficulties, alternative methodologies that are not based on spectral measurements have appeared such as estimating photometric redshifts \citep[e.g.,~][]{kaur17,kaur18,rajagopal20,rajagopal21}, finding associated galaxy groups \citep[e.g.,~][]{torres-zafra18}, or using machine learning techniques \citep[e.g.,~][]{coronado-blazquez23}. Another possibility is using a combination of data from ground-based imaging atmospheric Cherenkov telescopes and the LAT \citep{prandini10,yang10,acciari23}. This latter technique is based on the phenomenon that $\gamma$-ray photons coming from cosmological distances interact by pair-production with the extragalactic background light \citep[EBL, e.g.,~][]{hauser01,driver08,dwek13,driver21,saldana-lopez21}, leaving a signature in the $\gamma$-ray spectra that is dependent on the $\gamma$-ray photon energy and redshift \citep[e.g.,~][]{ebl12,hess_ebl13,dominguez15,biteau15,hess_ebl17,ebl18,magic_ebl19,veritas_ebl19,desai19,franceschini21,biteau22}. However, blazars tend to be rather variable sources, which makes combining often non-simultaneous data from different telescopes difficult and subject to systematic biases.

In this paper, we propose a methodology based solely on LAT data and EBL attenuation to set upper limits on the redshifts of blazars. This work is organized as follows. Section~\ref{sec:theo} includes a brief review on the theoretical aspects about $\gamma$-ray attenuation. In Section~\ref{sec:data}, we describe the $\gamma$-ray sample, data reduction, and methodology. This is followed by Section~\ref{sec:results}, where our results are shown and discussed. Finally, we summarize and conclude in Section~\ref{sec:summary}.

\section{Theoretical background on $\gamma$-ray attenuation}
\label{sec:theo}

We refer the reader to other detailed papers on the theoretical background on $\gamma$-ray attenuation such as \citet{dwek13} and \citet{finke22}, and briefly summarize the main aspects here.

There is a pair production interaction between $\gamma$ rays and EBL photons that produces an optical depth $\tau(E,z)$ that is analytically given by

\begin{equation}
\label{attenu}
\tau(E,z)=\int_{0}^{z} \Big(\frac{dl}{dz'}\Big) dz' \int_{0}^{2}d\mu \frac{\mu}{2}\int_{\varepsilon_{th}}^{\infty} d\varepsilon{'}\ \sigma_{\gamma\gamma}(\beta{'})n(\varepsilon{'},z'),
\end{equation}

\noindent where $E$ and $z$ are respectively the observed energy and redshift of the $\gamma$-ray source, $\sigma_{\gamma\gamma}$ is the photon-photon pair production cross section and $n(\varepsilon{'},z')$ is the proper number density of EBL photons with rest-frame energy $\varepsilon{'}$ at redshift $z'$ given by
\begin{equation}
    \label{num_density}
    \varepsilon{'}n(\varepsilon{'},z')=\frac{4\pi}{c\varepsilon{'}}\lambda I_{\lambda}(hc/\varepsilon{'},z').
\end{equation}

\noindent where $\lambda I_{\lambda}$ is the EBL spectral intensity. The cross section $\sigma_{\gamma\gamma}$ depends on the relative rest-frame energies of the $\gamma$-ray photon ($E'$), the rest mass energy of the electron $m_{e}c^2$, and the EBL photon energy ($\varepsilon{'}$) as
\begin{equation}
    \beta{'}=\frac{\varepsilon_{th}}{\varepsilon{'}},
\end{equation}
where $\varepsilon_{th}$ is the energy threshold for photon-photon pair production, which is given by
\begin{equation}
\label{threshold}
\varepsilon_{th}\equiv \frac{2m_{e}^2c^{4}}{E'\mu}
\end{equation}
and variable $\mu=(1-\cos\theta)$ connects the energy threshold to the angle of interaction $\theta$.

While $\gamma$-ray absorption due to the EBL does not have a fixed onset and strongly depends on redshift, a pragmatic choice of 10 GeV as a threshold can be made when considering blazars in the $z=2-3$ range. The observational effect is that this EBL attenuation leaves a characteristic signature on the $\gamma$-ray spectra of blazars \citep[e.g.,~][]{ebl12,ebl18}.


\section{Sample selection, data reduction, and methodology}\label{sec:data}
\subsection{Source selection}\label{sec:sources}

For our analysis, we select all blazars in the 4FGL-DR2 catalogue \citep{4FGL-DR2}. There are several definitions in use to distinguish BL Lacertae objects (BL Lacs) and flat spectrum radio quasars (FSRQs), but the most common one is as follows.  Blazars are classified as BL Lacs if the equivalent width of their emission lines is $<5$\ \AA\; and FSRQs otherwise \citep{marcha96,healey08}.
BL Lacs are expected to be the best targets for EBL studies since these are the sources exhibiting a hard $\gamma$-ray spectrum and unlikely to have an external photon field, e.g., broad line region (BLR), that may lead to additional $\gamma$-ray attenuation \citep[e.g.,~][]{dominguez15}. Although, \citet{ebl18} show that internal absorption is unlikely to be significant for FSRQs, a result confirmed by \citet{costamante18}. In the LAT catalogues there are also sources classified as blazar candidates of uncertain type (BCUs), which are objects that have a broad-band spectral energy distribution typical of blazars but do not have a spectral classification \citep{4LAC}.

Emissions from blazars can present variability producing changes of flux and modifications in the spectrum \citep[e.g.,~][]{penil20,penil22,oterosantos22}. These effects may lead to inconclusive results for information extracted from individual spectra using $\gamma$-ray attenuation \citep[see for instance the section on Variability in the Supplementary Material by][]{ebl18}, therefore in our approach we will distinguish between variable and non-variable blazars. This distinction is made using the \texttt{Variability\_Index} column in 4FGL-DR2 and using the condition lower or higher than 18.48 as explained by \citet{4FGL}. A higher value corresponds to a 99\% probability of the source being variable.

 Before making any calculations, we expected that our results would be significant only for BL Lacs and BCUs that are probably BL Lacs, since these sources typically have the strongest EBL signature. However we also ran the method on FSRQs as a sanity check. The initial number of sources in our sample was (non-variable/variable blazars) 621/510 BL Lacs, 866/446 BCUs, and 129/565 FSRQs.


\subsection{Data reduction and methodology} \label{sec:method}

For each source in our sample, we develop a LAT analysis as follows:
\begin{enumerate}
\item We perform an analysis in the 1 GeV to 2 TeV range without an EBL model, in order to determine the parameters that define the spectral function such as index and flux, also known as spectral parameters, of the sources around the source of interest. The spectral functions are either PowerLaw or LogParabola models depending on the best fit provided by 4FGL (see below for details).
\item We perform an analysis in the 1--10 GeV range without an EBL model, using the source spectral parameters (other than the source of interest) frozen to the values from step (i). The upper limit of 10 GeV is adopted considering the fact that the EBL absorption is negligible below this energy \citep{ebl12,ebl18}.
\item We extrapolate the spectrum from (ii) to 2 TeV, and include absorption with the EBL model, assuming a redshift of $z=0.01$, and record the log-likelihood value.  We repeat this step for all redshifts between $z=0.01$ and $z=3.00$ in 100 steps and save the log-likelihood value at each redshift.
\item We compare the log-likelihood values from step (iii) with the one from step (i) to create a TS profile.  The peak of this profile gives the most likely value for the redshift.  We take this redshift to be an upper limit. Because there can be a component in the spectral curvature of the sources that is not produced by EBL attenuation but which is intrinsic to their emission.
\end{enumerate}

A detailed description of this analysis is as follows. We perform a standard likelihood analysis on the {\it Fermi}-LAT data covering a given time period (first 14 years of telescope time), energy range (from 1~GeV to 2~TeV), and a selected region of interest (ROI) using \texttt{fermiPy} \citep{wood17}. The 4FGL catalogue of {\it Fermi}-LAT detected sources Data Release 2 \citep{4FGL,4FGL-DR2} is used to model the $\gamma$-ray sky including the latest interstellar emission model ({\it gll\_iem\_v07.fits}) and the standard template for the isotropic emission ({\it iso\_P8R3\_SOURCE\_V2\_v1.txt}). 
We have in the likelihood fitting all 4FGL-DR2 sources lying within the size of the circular region of interest (ROI) plus a radius $R$. The ROI size as well as $R$ depend on the minimum energy ($E_{\rm min}$) of the analysis. Since the point spread function of the {\it Fermi}-LAT improves considerably at 1 GeV compared to that at the usually adopted $E_{\rm min}=100$ MeV, we considered an ROI with radius of 7$^{\circ}$ and $R=3^{\circ}$. Spectral parameters associated with the sources lying outside the ROI are kept fixed, whereas those within the ROI are allowed to vary during the likelihood fitting. For each source, we use the same spectral model as adopted in the 4FGL-DR2 catalog. After the first round of the optimization, the ROI is scanned to search for unmodeled $\gamma$-ray sources by generating test statistic (TS) maps. The maximum likelihood TS is defined as TS = 2$\log$($L_{1}$-$L_{0}$), where $L_{0}$ and $L_{1}$ denote the likelihood values of the null hypothesis (i.e.,~no source at the position of interest) and that of the alternative hypothesis (i.e.,~the existence of a point source), respectively \citep{mattox96}. 
When an excess emission (TS$>25$) is found, it is added to the sky model following a power-law spectral model and, iteratively, a second set of TS maps is generated. A final likelihood fit is performed to optimize all the free parameters in the ROI once all excesses above the background are identified and inserted to the model to optimize the sky model as best as possible. This procedure is necessary since our data are taken in longer exposure than those from which the 4FGL-DR2 sky maps were constructed.

Next, we freeze the spectral parameters of all sources to their best-fitted values except for the source of interest. Then, we apply the EBL attenuation to the spectral model of the source of interest. For this purpose, we update the spectral model of the blazar either to {\it EBLAtten::PowerLaw2} or {\it EBLAtten::LogParabola} depending on whether the 4FGL-DR2 spectrum of the blazar reveals a significant curvature or not, which is given by the {\it Spectrum\_Type} column \citep[see][for details]{4FGL,4FGL-DR2}. In particular, the EBL attenuation is inserted in the spectral {\it PowerLaw2} or {\it LogParabola} models\footnote{\url{https://fermi.gsfc.nasa.gov/ssc/data/analysis/scitools/source_models.html}} as $\exp[-\tau(E,z)]$, where $\tau$ is the EBL optical depth as a function of the energy of the $\gamma$-ray photon and the redshift, $z$. In the first step, we use $\tau=0$, i.e., no EBL, and perform the likelihood fitting in the energy range 1$-$10 GeV. This exercise enables us to determine the intrinsic shape of the $\gamma$-ray spectrum, including any possible curvature. Then, we extrapolate the fitted spectrum to 2 TeV and allow the EBL optical depth to vary in the likelihood fit by changing the redshift value from $z=0.01$ to $z=3$ in 100 linear steps. In every step, the log-likelihood value is calculated thus permitting us to generate a likelihood profile as a function of redshift for each blazar in the sample.

\begin{figure}
	\includegraphics[width=\columnwidth]{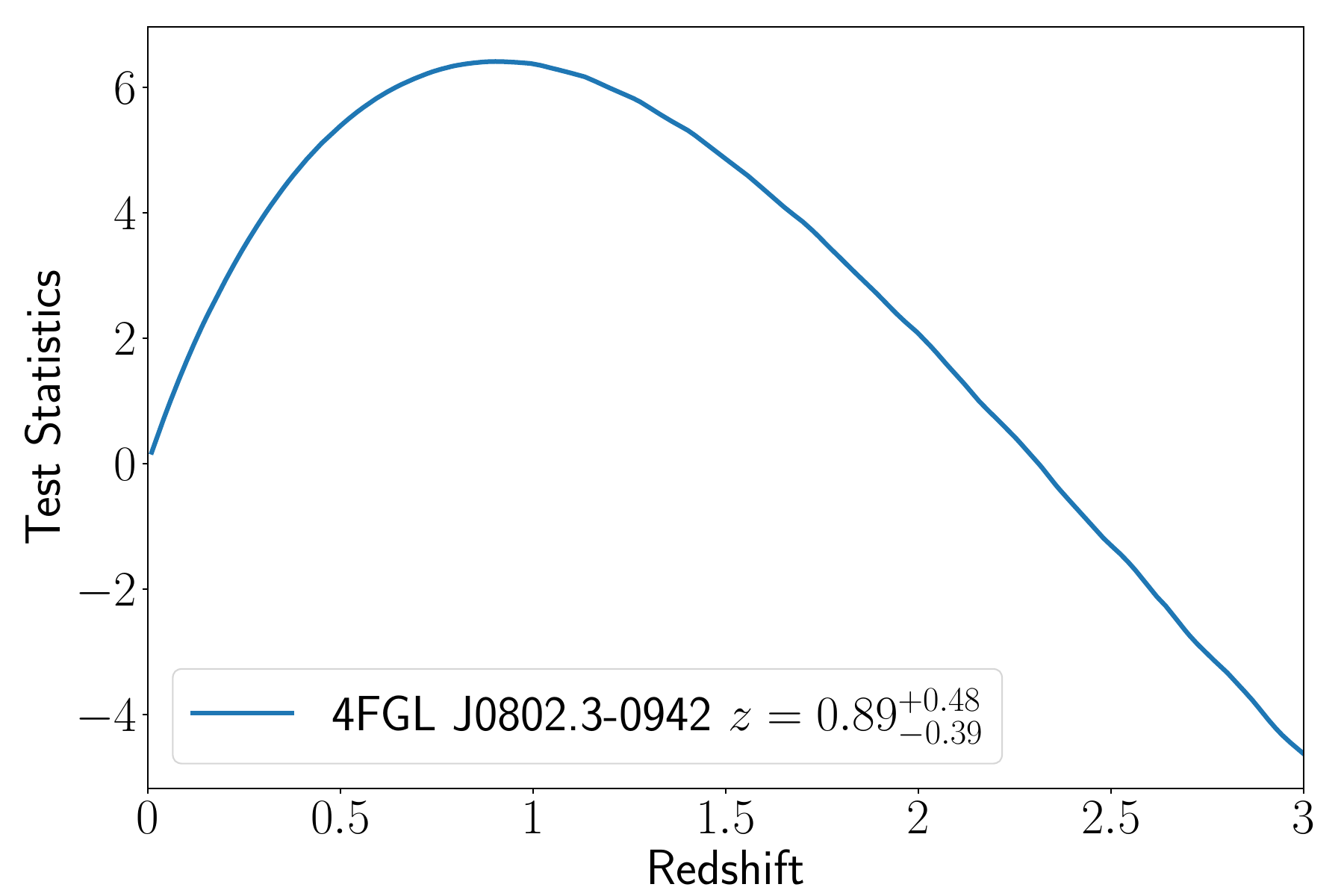}
    \caption{An example of one of the TS profiles. This is the case of 4FGL J0802.3-0942 with TS$_{peak}=38.3$. Note that negative TS, but not negative TS$_{peak}$, are possible at the highest redshifts and they mean that the alternate hypothesis of having a point source  with $\gamma$-ray spectrum attenuated by EBL at a given redshift is disfavoured w.r.t. to that of having no EBL. This may be an indication that the EBL model overestimates the attenuation at these redshifts.}
    \label{fig:TS}
\end{figure}

Later, we create a different TS profile by comparing this likelihood as a function of redshift with the likelihood from the null hypothesis of no EBL. Therefore, this TS is related to the EBL strength. The peak of the TS profile (TS$_{peak}$) corresponds to the maximum-likelihood value of the redshift of the target blazar, which, due to previous considerations on the possible intrinsic curvature of the spectra, should be interpreted as an upper limit. We remove from our results (1) all sources with TS$_{peak}<1$ since these correspond to noisy profiles and 2) those sources whose likelihood profiles do not show a peak in the 0.01$-$3 redshift range. We determine confidence limits from the TS profile. Figure~\ref{fig:TS} shows an example of one likelihood profile, in particular, for 4FGL J0802.3-0942.

 For estimating the impact on the redshifts due to the uncertainties of the intrinsic spectrum, we implement a Markov Chain Monte Carlo simulation for a subset of test blazars. For each blazar, we generate 100 random values from a Gaussian distribution centered on the best-fit intrinsic slope with a standard deviation equal to its uncertainty. We then rerun the LAT spectral analysis with these values fixed as the intrinsic spectrum. The resultant redshift distributions are consistent with our original results within the statistical uncertainties. For example, for J0540.5+5823, the redshift of TS$_{peak}$ obtained is $z$=1.0$\pm$0.3, while from our simulation, we find a redshift distribution whose mean is 1.1 and standard deviation 0.4. Similar consistency is observed for the other blazars we tested, indicating that the impact of the statistical uncertainties on the redshift limits is minimal. These results hold at a 68\% confidence level, reinforcing the robustness of our original findings.

\section{Results and discussion}\label{sec:results}

\subsection{Redshift upper limits}
There are 344/331 non-variable/variable sources that satisfy the criteria that are given at the end of Section~\ref{sec:method}. From this sample, there are 5/40 non-variable/variable FSRQs, all of them, with redshifts in the 4LAC catalogue \citep{4LAC}, except 4FGL J1532.7-1319. However, we note that this source is classified as BCU in the SIMBAD database\footnote{\url{https://cds.u-strasbg.fr/}}. 

\begin{figure*}
	\includegraphics[width=\columnwidth]{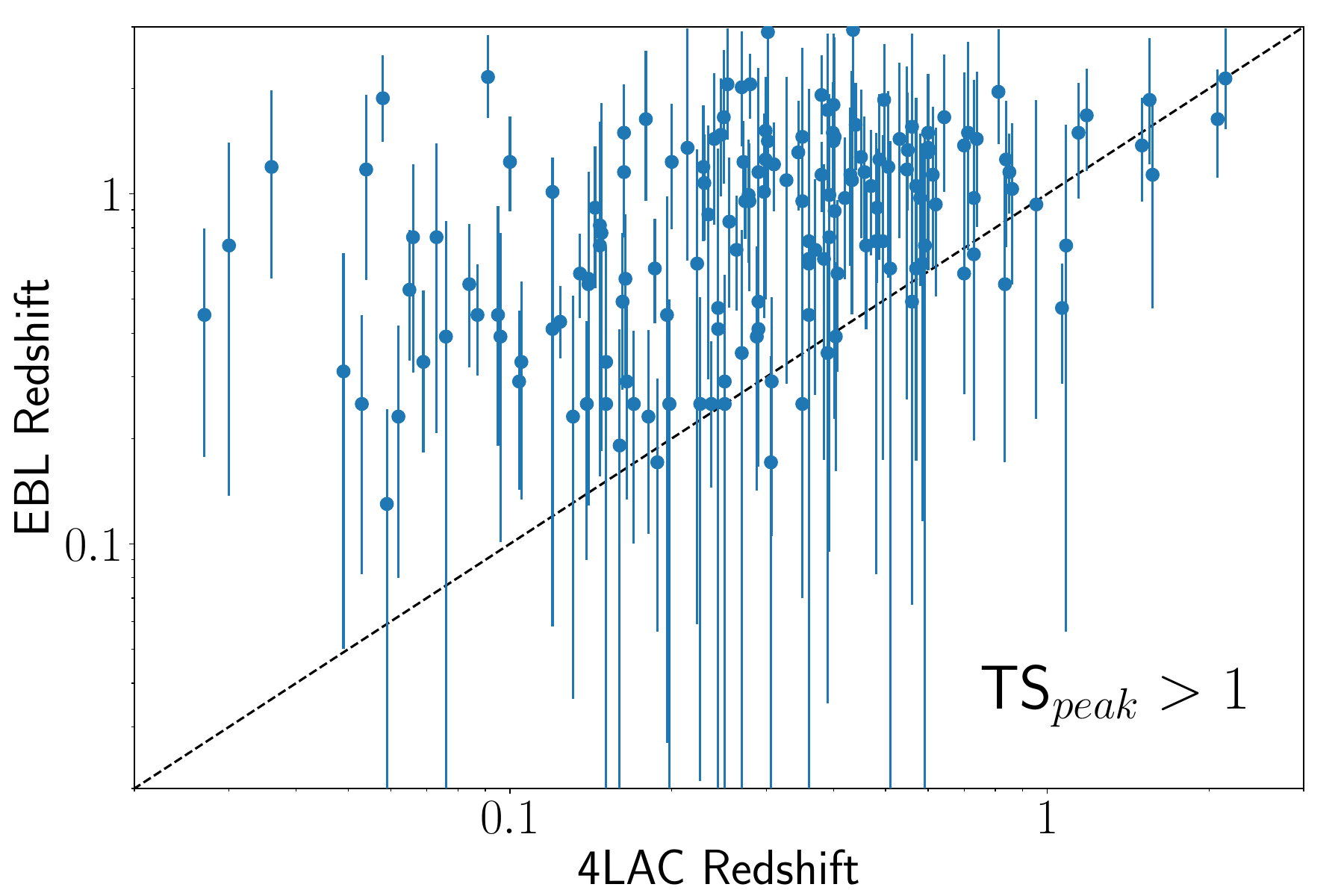}
	\includegraphics[width=\columnwidth]{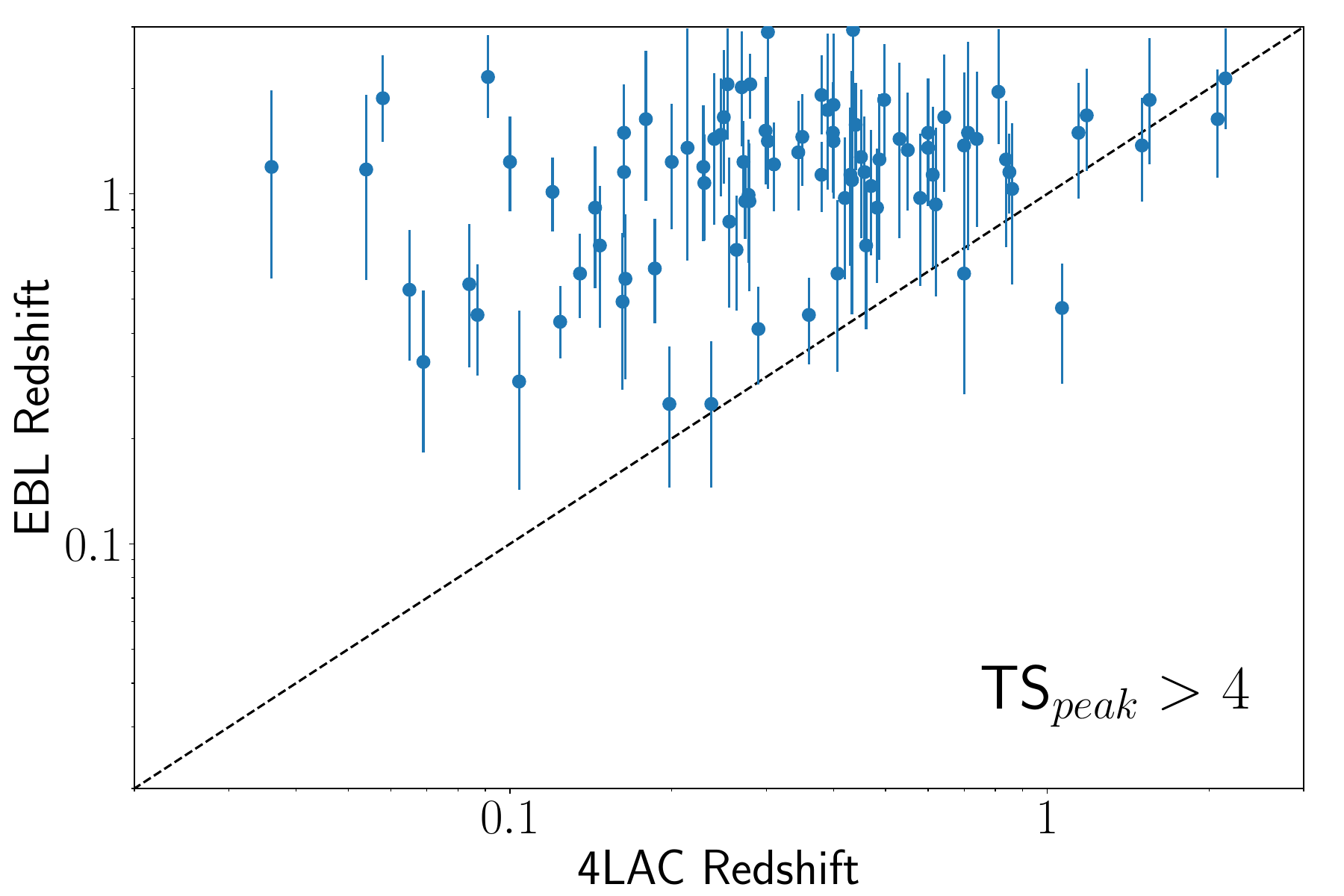}
    \caption{Redshift from our EBL methodology for non-variable blazars ({\it Left panel}) for the 166 sources with redshift in 4LAC and TS$_{peak}\geq 1$ and ({\it Right panel}) the 91 sources with redshift in 4LAC and TS$_{peak}\geq 4$. The uncertainties at lower redshifts, $z<0.5$, can be larger than the ones at higher redshifts, $z\sim 1$, because at the lower redshifts, the EBL affects energies larger than 200--300~GeV where the LAT's effective area is smaller.}
    \label{fig:check}
\end{figure*}

\begin{figure*}
	\includegraphics[width=\columnwidth]{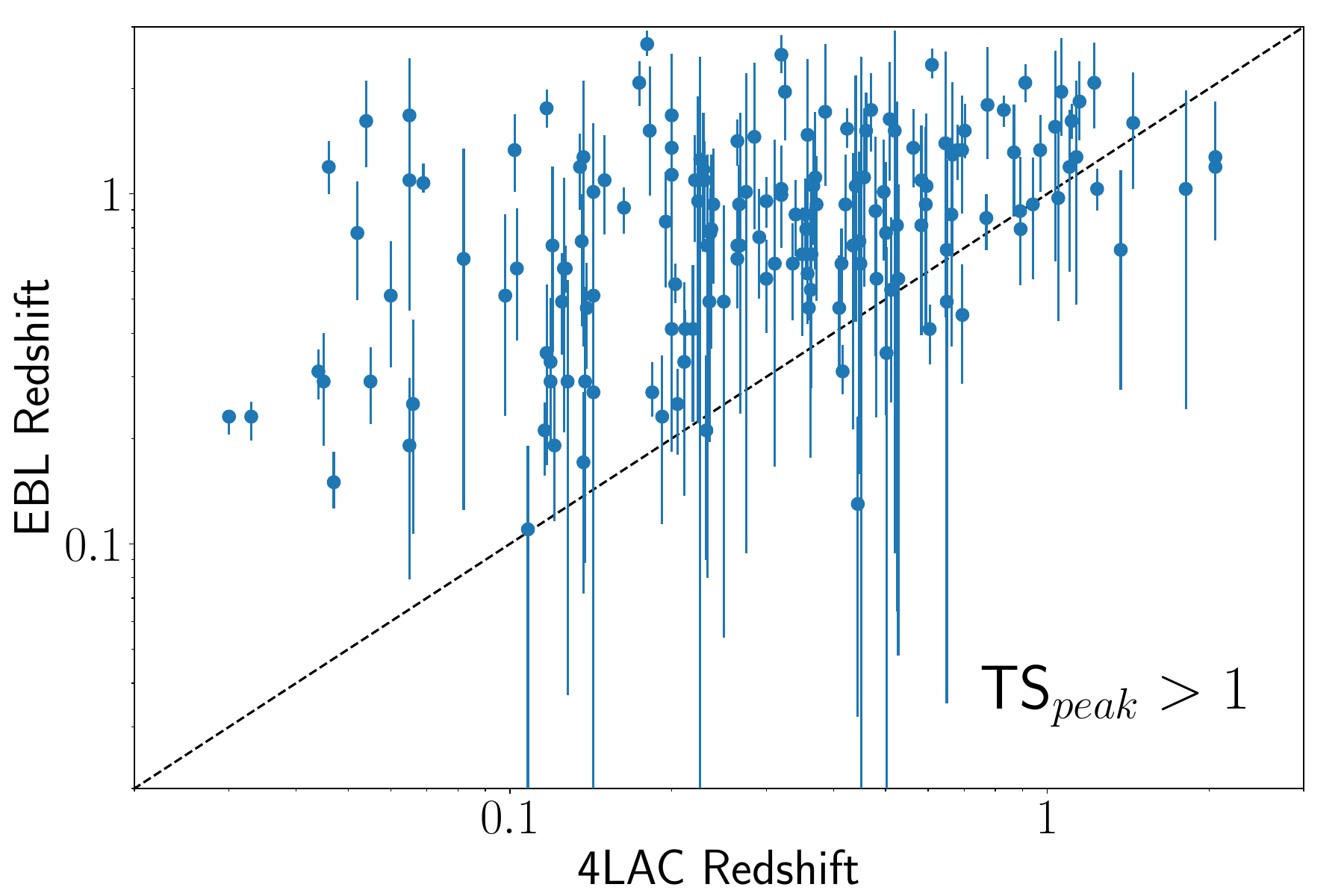}
	\includegraphics[width=\columnwidth]{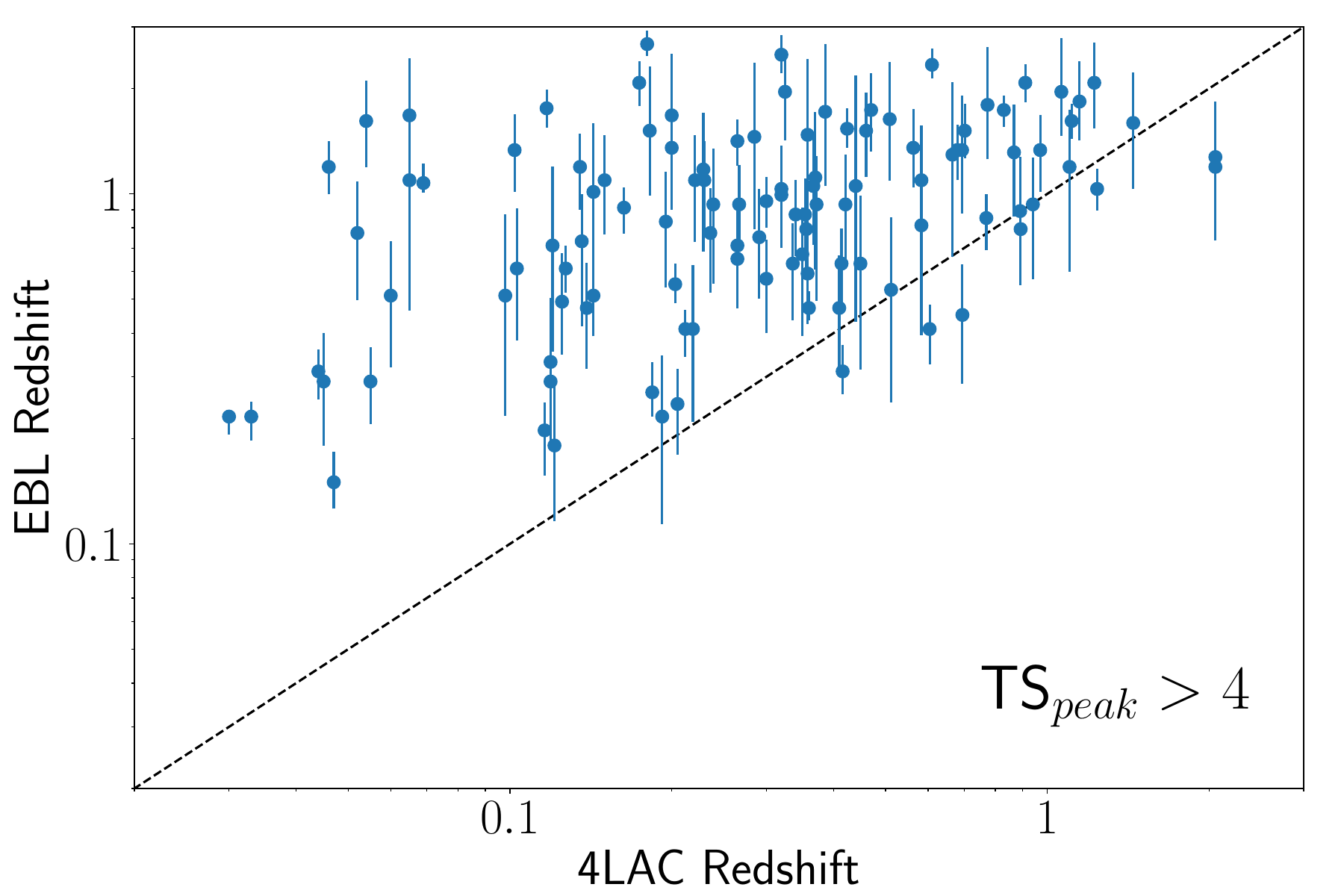}
    \caption{Redshift from our EBL methodology for variable blazars ({\it Left panel}) for the 162 sources with redshift in 4LAC and TS$_{peak}\geq 1$ and ({\it Right panel}) the 111 sources with redshift in 4LAC and TS$_{peak}\geq 4$. See comment in the previous figure about the uncertainties.}
    \label{fig:check2}
\end{figure*}

Then, we have 338/293 non-variable/variable sources (BL Lacs and BCUs), from which 166/162 have a redshift in the 4LAC catalogue and 172/130 do not have a redshift. As a sanity check we plot in Figure~\ref{fig:check}, the redshift obtained from our EBL-attenuation methodology versus the one in 4LAC for the 166 sources with redshift and TS$_{peak}\geq 1$, and also for the 91 sources with redshift and TS$_{peak}\geq 4$. Figure~\ref{fig:check2} shows similar information for the variable blazars, 162 with TS$_{peak}\geq 1$ and 111 with TS$_{peak}\geq 4$. We can see empirically for both sub-samples that the redshift obtained from our methodology can be considered an upper limit for the redshift of the sources in both cases. This is expected since there can be a component in the spectral curvature of the sources that is not produced by EBL attenuation but which is intrinsic to their emission. We tested whether our results change when they are extracted from different periods for the variable sources, concluding that the redshift upper limit generally varies significantly. However in the vast majority of cases, from the comparison with known redshifts, the result is still an upper limit. Furthermore, the empirical test shown in Figure~\ref{fig:check2} also supports the validity of our upper limits.


For the non-variable blazars, and the case with TS$_{peak}\geq 1$ (and also TS$_{peak}\geq 4$), there is only one source that is away from the one to one line by more than $1\sigma$, which are also within the statistical expectations. This source is Ton~116 ($z_{4LAC}=1.065$ vs. $z_{EBL}<0.632$ at 84\% C.L., TS$_{peak}=5.69$). The redshift of Ton~116 given by SIMBAD comes from \citet{sdss09}. We could not see any feature leading to that redshift and we think that it could have been automatically assigned by the line search pipeline, probably with low significance. Yet there is a lower limit by \citet{paiano17} at $z>0.48$ that is compatible with our upper limit of $z_{EBL}<0.632$ (84\% C.L.).

In the case of the variable blazars there are eight blazars that are away from the one to one line more than 1$\sigma$ (although all within 2$\sigma$), also in agreement with statistical expectations. These sources are:

\begin{itemize}

\item 3C 66A ($z_{4LAC}=0.444$ vs. $z_{EBL}<0.150$ at 84\% C.L., TS$_{peak}=1.66$). The $z_{4LAC}=0.444$ value comes from very weak lines reported in \citet{Miller1978}; \citet{Kinney1991}, and \citet{lanzetta93}.  Based on the marginal detection of the host galaxy by \citet{wurtz96}, \citet{finke08_redshift} estimated the redshift to be $z=0.321$.  However, \citet{Stadnik2014} were not able to detect the host galaxy and, based on this non-detection, gave a lower limit of $z>0.42$. In SIMBAD, the reported value for the redshift of this source is $z_{SIMBAD} = 0.340$, using the HI column density \citep{Neeleman2016}. Based on a comparison of the LAT and MAGIC spectrum, and assuming EBL attenuation, \citet{aleksic11} estimated the redshift $z<0.68$. Recently, \citet{Paiano2017a} observed the optical spectra of 3C 66A in a spectroscopic campaign carried out at the 10 m Gran Telescopio Canarias, leading to inconclusive results. To this date, the redshift of this source remains a subject of debate.

\item 1ES 0502+675 ($z_{4LAC}=0.416$ vs. $z_{EBL}<0.331$ at 84\% C.L., TS$_{peak}=54.39$). As for the previous case, the redshift reported in SIMBAD is different from the value that appears in the 4LAC. In 4LAC, the redshift value $z_{4LAC}=0.416$ is taken from \citet{Landt2002} , while in SIMBAD the value $z_{SIMBAD}=0.314$ is selected from \citet{Sbarufatti2000}, that is compatible with the upper limit here derived. However, these studies rely on imaging methods, since there is no optical spectrum available for this source.

\item TXS~0628$-$240 ($z_{4LAC}=1.238$ vs. $z_{EBL}<1.124$ at 84\% C.L., TS$_{peak}=45.70$). The redshift comes from \citet{landt12}, which actually gives a lower limit based on a clear Mg II absorption line. This $z>1.238$ limit is similar to our upper limit of $z_{EBL}<1.124$ (84\% C.L.), likely indicating that the Mg II line actually comes from the BL Lac host galaxy instead of an intercepting gas cloud.

\item NVSS J090226+205045 ($z_{4LAC}=2.055$ vs. $z_{EBL}<1.274$ at 84\% C.L., TS$_{peak}=5.44$).The 4LAC redshift is taken from the automatic line finding pipeline given by the SDSS DR6 \citep{Richards2009}, which can be biased for faint lines as are typical in BL Lacs. However, there are no emission/absorption lines in the spectra that allow its confirmation. A more recent upper limit for the redshift of this object is available in the literature: $z < 1.21$ given by \cite{Rau2012} using UV-to-NIR photometry.

\item PKS 0903-57 ($z_{4LAC}=0.695$ vs. $z_{EBL}<0.451$ at 84\% C.L., TS$_{peak}=7.82$). This value is selected from optical spectroscopic observations performed by \citet{Thompson1990}. However, these authors select the position of the source by its radio emission, and they mention that the optical counterpart corresponding to the radio emission is a star, so the optical spectra they took corresponds to a source 4 arcsec west of the position of the radio emission, a Seyfert I. Therefore, it would be necessary to re-observe the optical spectra for this object, to see whether the optical counterpart observed to determine the value $z=0.695$ is correct or is a different source.

\item 87GB 105148.6+222705 ($z_{4LAC}=2.055$ vs. $z_{EBL}<1.274$ at 84\% C.L., TS$_{peak}=6.71$). As in the previous case, the 4LAC redshift value is taken from the SDSS DR6 \citep{Richards2009}, while this value is assigned automatically with poor significance. \cite{Rau2012} gave an upper limit of $z < 1.36$, and there is a more recent value given automatically by the SDSS DR13\footnote{\url{https://www.sdss.org/dr13/data\_access/bulk/}}  for this object: $z = 0.63$, both compatible with the upper limit presented here.


\item PKS 1424+240 ($z_{4LAC}=0.604$ vs. $z_{EBL}<0.431$ at 84\% C.L., TS$_{peak}=25.17$). This redshift is obtained by \citet{paiano17b} from some weak emission lines. \citet{Rovero2016} identifies this BL Lac as a member of a galaxy cluster at $z=0.601$.

\item  MG3 J225517+2409 ($z_{4LAC}=1.370$ vs. $z_{EBL}<0.752$ at 84\% C.L., TS$_{peak}=2.77$). The 4LAC redshift value is from \citet{Albareti2017}, but we did not identify any feature leading to that redshift and we think that it could have been automatically assigned, probably with small significance. 

\end{itemize}

Since the upper limit interpretation is robust for non-variable as well as for variable sources, we do not distinguish between them in the following discussions. Table~\ref{tab1} lists the 303 blazars that do not have a redshift in 4LAC and whose likelihood profile peaks at $z<3$. We stress that the 125 sources with 1 $\leq$ TS$_{peak}$ < 4 correspond to a significance lower than 2$\sigma$, which according to Figure~\ref{fig:check} seem to be compatible with upper limits as well. At any rate, although we give these low significance results (TS$_{peak}< 4$) in Table~\ref{tab1}, we suggest the user be prudent about using them.

\begin{figure}
	\includegraphics[width=\columnwidth]{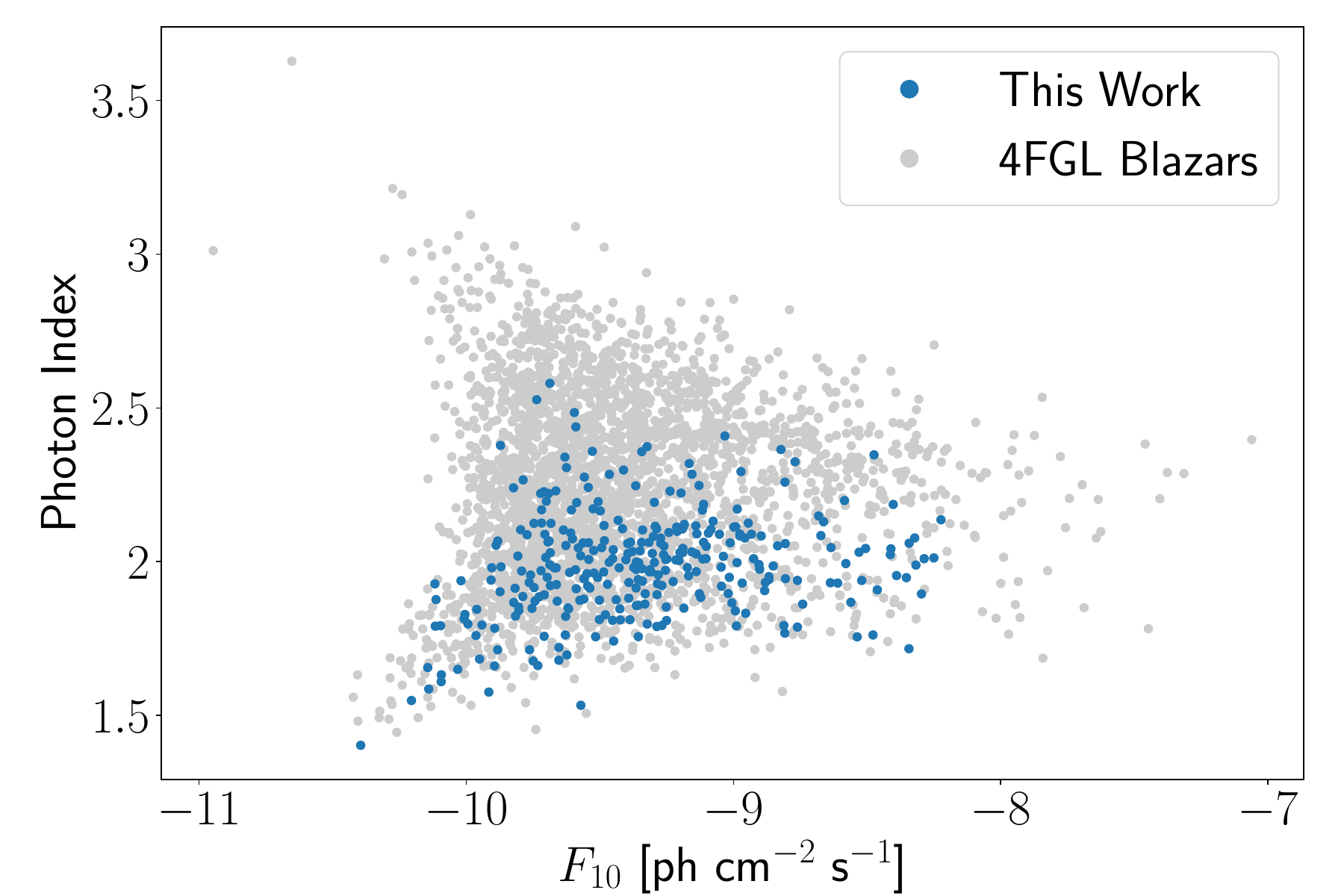}
    \caption{Photon index versus integrated flux at energies above 10 GeV from the initial sample (grey symbols) and the 303 blazars for which we find a redshift upper limit (blue symbols).}
    \label{fig:sel}
\end{figure}

Figure~\ref{fig:sel} shows the 303 sources listed in Table~\ref{tab1} in comparison with our initial source selection. It is clear that these blazars lie in the lower region of the initial sample cloud, where, as expected, they tend to have a lower photon index, i.e.,~hard spectrum.

\begin{figure*}
	\includegraphics[width=2\columnwidth]{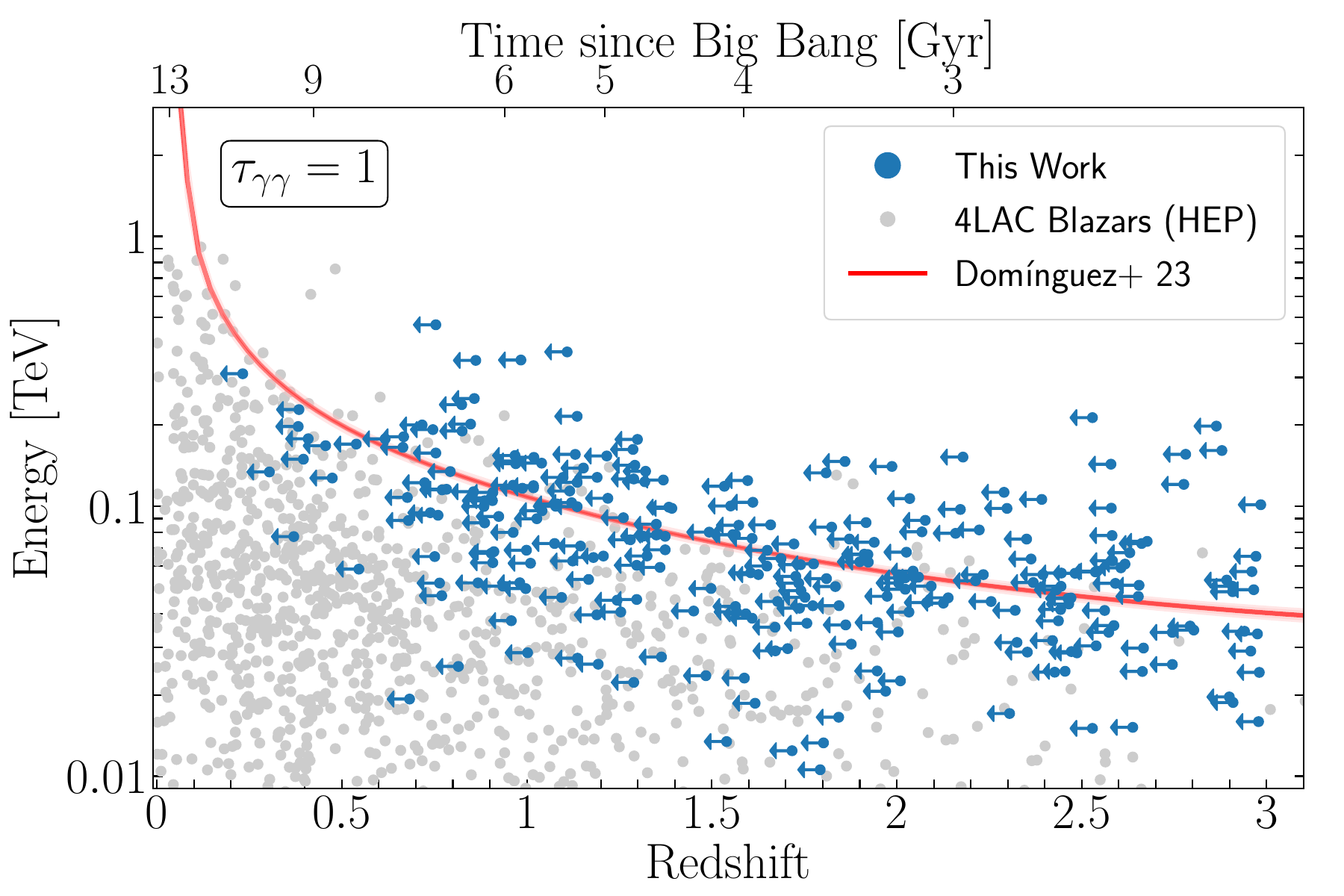}
    \caption{Highest energy photons of 4LAC blazars (grey circles) as a function of redshift, the cosmic $\gamma$-ray horizon from \citet[][red band]{dominguez23} based on the EBL model by \citet{saldana-lopez21}, and the upper limits at 84\% C.L. from this work (blue circles). We find upper limits for the redshifts of 303 blazars.}
    \label{fig:cgrh}
\end{figure*}

 In Figure~\ref{fig:cgrh}, we show the 303 blazars using the redshift upper limit at 84\% C.L. along the cosmic $\gamma$-ray horizon (CGRH), this is the energy as a function of redshift for which $\tau=1$. This implies CGRH divides the universe in two different regions for which the surviving flux is larger or smaller than $\exp(-1)\sim $37\%. The upper limits that lie around the most opaque regions, such as at $z\sim 2.5$ and energies of approximately 200~GeV are poorly constrained.

In order to estimate the uncertainties introduced by the EBL, we also run our pipeline using the lower uncertainties of the optical depths given by \citet{dominguez23} based on the model by \citet{saldana-lopez21}. These lower uncertainties will translate into larger values of the redshift where the TS profile peaks, which goes from smaller shifts in TS$_{peak}$ for the lower redshifts and larger for the higher redshifts. This effect is a consequence of having better constrained the EBL at the lower than at the higher redshifts. The increase in the redshifts where TS$_{peak}$ is, in redshift bins of 0.5 from $z=0$ to $z=3$, is on average approximately 13\%, 16\%, 21\%, 28\%, 26\% and 30\%, respectively. We stress that the redshifts upper limits obtained for other models such as \citet{finke10}, \citet{dominguez11a}, \citet{franceschini17} and \citet{finke22} are within these uncertainties. Any larger intensity EBL will not affect the redshifts upper limits because a more opaque universe leads to lower redshifts for the TS$_{peak}$. Furthermore, we tested that TS$_{peak}$ tend to be larger for \citet{dominguez23} than for \citet{dominguez11a} indicating that the former model produces better fits, as expected since it is based on deeper and extended multiwavelength data.

\subsection{Application case: detectability predictions for the Cherenkov Telescope Array}
The redshift upper limits are used to compute the expected detection significance of the 303 sources listed in Table~\ref{tab1} with the Cherenkov Telescope Array \citep[CTA,][]{2019scta.book.....C} using a total exposure of 20 hours, a reasonable exploratory time for IACTs. For the estimate of the source detectability, we use the CTA instrument response functions\footnote{\url{https://zenodo.org/record/5499840}} (IRFs) for the Alpha configuration with average azimuth angle, which are available for three different zenith angles (i.e.,~20 deg, 40 deg and 60 deg), both for the northern and southern arrays. The CTA-North and CTA-South IRFs are used for the sources with positive and negative declination, respectively. We assume observations around culmination time to select the most adequate IRF configuration concerning the zenith angle. We have generated point-like differential sensitivity curves for a total exposure time of 20 hours by scaling the IRFs corresponding to an exposure of 5 hours utilizing \texttt{gammapy} \citep{2017ICRC...35..766D,acero_fabio_2022_7311399}. As products from this calculation we obtained the differential flux, number of excess events, and number of background events that would generate a $5 \sigma$ signal in each energy bin. This is then used to obtain the expected number of excess events for each considered source, which is computed by scaling linearly the number of excess events necessary to get a detection significance of $5\sigma $ with the ratio of the differential flux. This is the differential flux of the sensitivity curve for each energy bin. The obtained number of excess and background events are used to estimate the statistical detection significance of the blazars using the function \texttt{WStatCountsStatistic} of \texttt{gammapy}, which applies the formula derived by \citet{1983ApJ...272..317L}. The energy threshold in each case is selected as the energy of the lower bin of the corresponding differential sensitivity curve, i.e. the energy threshold is considered larger for sources that can only be observed with a zenith angle $>50$~deg and also for sources observed with CTA-South. 

For the source's spectral shape, we consider the result from the LAT data analysis at the redshift upper limit, which we stress is a PowerLaw or LogParabola including the EBL exponential attenuation. This spectral shape is extrapolated to TeV energies considering the EBL absorption following the optical depths by \citet{dominguez23} based on the recent model by \citet{saldana-lopez21}, consistent with our LAT data reduction. The significance is estimated using as the redshift of the source, the upper limit values at 84\% C.L. from Table~\ref{tab1}; this is the redshift of the TS$_{peak}$ plus upper uncertainties. For these calculations we use the best zenith angle for the observation either from CTA-North or CTA-South. This results in a total of 21 sources with an expected significance of $> 5 \sigma$. We stress that this refers to detections in the average/quiescent state of the source derived from the average flux over 14 years of {\it Fermi}-LAT telescope time. Note that considering only the redshift of the TS$_{peak}$, which can be a reasonable approach according to Figure~\ref{fig:check}, there are 41 sources with a significance of $> 5 \sigma$. These additional 20 sources are J0143.7-5846, J0322.0+2335, J0333.9+6537, J0334.2-3725, J0338.5+1302, J0338.9-2848, J0540.5+5823, J0600.3+1244, J0709.2-1527, J0812.0+0237, J0826.4-6404, J1240.4-7148, J1253.2+5301, J1454.4+5124, J1518.0-2731, J1553.5-3118, J1610.7-6648, J1704.5-0527, J2104.3-0212 and J2247.8+4413. We stress that, since we use redshift upper limits in our computation, these CTA detection significances can be considered lower limits.  If the sources were closer, the EBL attenuation would be less, so the very-high-energy emission ($E>50$~GeV) from the sources would be brighter, and they would be easier for CTA to detect.

\section{Summary and conclusions}\label{sec:summary}
We have calculated redshift upper limits for 303 $\gamma$-ray blazars using LAT data and EBL attenuation. The majority of these blazars are classified as BL Lacs. The redshifts derived with our methodology can be useful for studying the evolution of blazars or planning observations with other telescopes, for instance, with imaging atmospheric Cherenkov telescopes such as MAGIC, VERITAS, H.E.S.S and CTA. We find that 21 blazars, for which we have found a redshift upper limit, can be detected in their average/quiescent state with CTA in 20h.

\section*{Acknowledgements}

The \textit{Fermi}-LAT Collaboration acknowledges generous ongoing support
from a number of agencies and institutes that have supported both the
development and the operation of the LAT as well as scientific data analysis.
These include the National Aeronautics and Space Administration and the
Department of Energy in the United States, the Commissariat \`a l'Energie Atomique and the Centre National de la Recherche Scientifique / Institut National de Physique Nucl\'eaire et de Physique des Particules in France, the Agenzia Spaziale Italiana and the Istituto Nazionale di Fisica Nucleare in Italy, the Ministry of Education,Culture, Sports, Science and Technology (MEXT), High Energy Accelerator Research Organization (KEK) and Japan Aerospace Exploration Agency (JAXA) in Japan, and the K.~A.~Wallenberg Foundation, the Swedish Research Council and the Swedish National Space Board in Sweden.

Additional support for science analysis during the operations phase is gratefully acknowledged from the Istituto Nazionale di Astrofisica in Italy and the Centre National d'\'Etudes Spatiales in France. This work was performed in part under DOE Contract DE-AC02-76SF00515.

This research has made use of the CTA instrument response functions provided by the CTA Consortium and Observatory, see \url{https://www.ctao-observatory.org/science/cta-performance/} \citep[version prod5 v0.1;][]{cherenkov_telescope_array_observatory_2021_5499840} for more details.

We thank Daniel Nieto for providing computational resources, David Paneque for helpful comments, and Jonathan Biteau for a constructive review. A.D. is thankful for the support of the Ram{\'o}n y Cajal program from the Spanish MINECO, Proyecto PID2021-126536OA-I00 funded by MCIN / AEI / 10.13039/501100011033, and Proyecto PR44/21‐29915 funded by the Santander Bank and Universidad Complutense de Madrid.  J.F.\ is partially supported by NASA through contract S-15633Y. M.L. is funded by MCIN through grant PRE2020-093502. J.L.C. is supported by MCIN project PID2019-104114RB-C32. J.F. is partially supported by NASA through contract S-15633Y.
\section*{Data Availability}

The results of this study are publicly available at \url{https://www.ucm.es/blazars/zebl} and by request to the authors.



\bibliographystyle{mnras}
\bibliography{redshifts_lat}



\begin{table*}
\begin{center}
\begin{tabular}{lccccrcc}
\hline
4FGL Source Name   &   Redshift Upper Limit & TS$_{peak}$ & HEP (GeV) & Type & Association & TeV & CTA ($\sigma$)\\ 
(1) & (2) & (3) & (4) & (5) & (6) & (7) & (8) \\ 
\hline
J0001.2$-$0747 & $1.19_{-0.61}^{+0.72}$ & 5.1 & 86 & bll & PMN J0001-0746 & N & 0.7 \\
J0001.6$-$4156 & $0.27_{-0.25}^{+0.97}$ & 1.2 & 44 & bcu & 1RXS J000135.5-415519 & N & 1.2 \\
J0002.1$-$6728 & $1.66_{-0.59}^{+0.71}$ & 9.8 & 52 & bcu & SUMSS J000215-672653 & N & 0.1 \\
J0003.1$-$5248 & $0.51_{-0.40}^{+0.41}$ & 1.7 & 91 & bcu & RBS 0006 & N & 1.1 \\
J0009.3+5030 & $0.77_{-0.30}^{+0.30}$ & 4.4 & 72 & bll & NVSS J000922+503028 & N & 3.0 \\
J0014.7+5801 & $1.96_{-0.58}^{+0.92}$ & 47.4 & 160 & bll & 1RXS J001442.2+580201 & N & 0.6 \\
J0014.8+6118 & $2.12_{-0.66}^{+0.82}$ & 16.27 & 34 & bcu & 4C +60.01 & N & 0.6 \\
J0015.6+5551 & $0.27_{-0.14}^{+0.18}$ & 3.8 & 167 & bll & GB6 J0015+5551 & N & 3.2 \\
J0017.0$-$0649 & $1.27_{-1.24}^{+1.38}$ & 1.1 & 46 & bcu & PMN J0017-0650 & N & 0.1 \\
J0019.3$-$8152 & $1.25_{-0.57}^{+0.56}$ & 4.8 & 53 & bll & PMN J0019-8152 & N & 0.0 \\
J0019.6+2022 & $1.07_{-0.94}^{+1.56}$ & 1.3 & -- & bll & PKS 0017+200 & N & 0.2 \\
J0021.0+0322 & $2.94_{-1.29}^{+0.02}$ & 12.3 & -- & bcu & 2MASS J00205023+0323578 & N & 0.3 \\
J0021.5$-$2552 & $1.19_{-0.72}^{+0.90}$ & 2.7 & 50 & bll & CRATES J002132.55-255049.3 & N & 0.3 \\
J0022.5+0608 & $1.33_{-0.47}^{+0.53}$ & 7.0 & 36 & bll & PKS 0019+058 & N & 1.7 \\
J0026.6$-$4600 & $0.61_{-0.49}^{+0.53}$ & 1.6 & 71 & bll & 1RXS J002636.3-460101 & N & 0.6 \\
J0031.3+0726 & $0.99_{-0.89}^{+1.06}$ & 1.3 & 54 & bll & NVSS J003119+072456 & N & 0.2 \\
J0035.9+5950 & $1.03_{-0.10}^{+0.08}$ & 739.8 & 372 & bll & 1ES 0033+595 & Y & 38.0 \\
J0040.3+4050 & $0.39_{-0.29}^{+1.14}$ & 2.4 & 118 & bll & B3 0037+405 & N & 0.8 \\
J0045.7+1217 & $0.67_{-0.29}^{+0.34}$ & 6.4 & 61 & bll & GB6 J0045+1217 & N & 3.0 \\
J0052.9$-$6644 & $1.76_{-0.73}^{+1.05}$ & 8.4 & -- & bcu & PMN J0052-6641 & N & 0.0 \\
J0115.6+0356 & $1.17_{-0.59}^{+0.56}$ & 4.1 & 51 & bll & PMN J0115+0356 & N & 1.4 \\
J0116.0$-$2745 & $0.99_{-0.66}^{+0.75}$ & 2.4 & 48 & bll & 1RXS J011555.6-274428 & N & 0.3 \\
J0116.2$-$6153 & $0.85_{-0.60}^{+0.64}$ & 2.1 & 80 & bll & SUMSS J011619-615343 & N & 0.1 \\
J0120.4$-$2701 & $0.79_{-0.19}^{+0.20}$ & 21.8 & 116 & bll & PKS 0118-272 & N & 12.3 \\
J0127.2+0324 & $1.09_{-0.59}^{+0.73}$ & 4.1 & 83 & bll & NVSS J012713+032259 & N & 0.6 \\
J0135.1+0255 & $0.87_{-0.49}^{+0.69}$ & 5.2 & 78 & bcu & 1RXS J013506.7+025558 & N & 0.5 \\
J0136.5+3906 & $0.73_{-0.12}^{+0.13}$ & 35.1 & 250 & bll & B3 0133+388 & Y & 21.5 \\
J0138.0+2247 & $0.47_{-0.42}^{+0.44}$ & 1.3 & 67 & bll & GB6 J0138+2248 & N & 1.2 \\
J0140.6+8736 & $1.07_{-0.79}^{+1.38}$ & 2.1 & 24 & bcu & WN B0126.6+8722 & N & 0.0 \\
J0143.7$-$5846 & $0.77_{-0.18}^{+0.20}$ & 25.2 & 154 & bll & SUMSS J014347-584550 & N & 4.3 \\
J0144.6+2705 & $2.38_{-0.39}^{+0.53}$ & 41.0 & 48 & bll & TXS 0141+268 & N & 2.2 \\
J0148.2+5201 & $0.41_{-0.26}^{+0.27}$ & 2.5 & 107 & bll & GB6 J0148+5202 & N & 2.0 \\
J0150.6$-$5448 & $0.61_{-0.56}^{+0.84}$ & 1.2 & 40 & bcu & PMN J0150-5450 & N & 0.1 \\
J0158.5$-$3932 & $0.59_{-0.49}^{+0.60}$ & 1.5 & 39 & bll & PMN J0158-3932 & N & 0.8 \\
J0159.7$-$2740 & $0.77_{-0.72}^{+0.59}$ & 1.1 & 27 & bll & PMN J0159-2739 & N & 0.5 \\
J0201.1$-$4347 & $0.91_{-0.81}^{+1.06}$ & 1.3 & 20 & bcu & GALEXASC J020110.83-434654.8 & N & 0.1 \\
J0206.8$-$5744 & $0.35_{-0.34}^{+0.46}$ & 1.0 & -- & bcu & SUMSS J020640-574948 & N & 0.7 \\
J0208.3$-$6838 & $1.44_{-0.72}^{+1.11}$ & 5.5 & 57 & bll & PKS 0206-688 & N & 0.0 \\
J0209.3+4449 & $0.57_{-0.44}^{+0.52}$ & 2.2 & 45 & bll & 1RXS J020917.6+444951 & N & 1.3 \\
J0215.3+7555 & $0.45_{-0.15}^{+2.13}$ & 14.3 & 143 & bcu & WN B0210.3+7540 & N & 0.4 \\
J0223.0+6821 & $1.31_{-0.59}^{+0.73}$ & 6.5 & 57 & bll & NVSS J022304+682154 & N & 0.6 \\
J0224.0$-$7941 & $1.25_{-1.24}^{+1.28}$ & 1.0 & 30 & bll & PMN J0223-7940 & N & 0.0 \\
J0226.5$-$4441 & $1.68_{-0.65}^{+1.10}$ & 11.7 & 120 & bll & RBS 0318 & N & 0.3 \\
J0233.9+8041 & $1.21_{-0.39}^{+0.51}$ & 20.0 & 72 & bcu & 1RXS J023428.6+804341 & N & 0.2 \\
J0240.8$-$3401 & $0.97_{-0.79}^{+1.11}$ & 1.6 & 43 & bcu & NVSS J024047-340018 & N & 0.1 \\
J0241.3+6543 & $0.51_{-0.40}^{+0.40}$ & 1.7 & 104 & bll & TXS 0237+655 & N & 1.5 \\
J0245.1$-$0257 & $1.01_{-0.70}^{+0.74}$ & 2.0 & 45 & bll & PMN J0245-0255 & N & 0.3 \\
J0250.7+5630 & $1.68_{-0.71}^{+0.77}$ & 5.7 & 50 & bcu & RX J0250.7+5629 & N & 0.6 \\
J0259.3+5453 & $1.13_{-0.76}^{+0.84}$ & 2.6 & 46 & bcu & GB6 J0259+5451 & N & 0.3 \\
J0303.6+4716 & $0.63_{-0.40}^{+0.37}$ & 2.1 & 68 & bll & 4C +47.08 & N & 2.4 \\
J0307.4+4915 & $1.44_{-0.47}^{+0.50}$ & 9.0 & 73 & bll & GB6 J0307+4915 & N & 1.2 \\
J0310.6$-$5017 & $0.87_{-0.41}^{+0.42}$ & 1.7 & 60 & bll & 1RXS J031036.0-501615 & N & 1.1 \\
J0314.3$-$5103 & $1.44_{-0.60}^{+0.70}$ & 7.5 & 45 & bll & PMN J0314-5104 & N & 0.4 \\
J0315.4$-$2643 & $0.57_{-0.44}^{+0.56}$ & 1.8 & 27 & bcu & NVSS J031527-264400 & N & 0.7 \\
J0316.2$-$6437 & $0.59_{-0.23}^{+0.26}$ & 9.7 & 201 & bll & SUMSS J031614-643732 & N & 2.6 \\
J0318.7+2135 & $1.38_{-0.49}^{+0.70}$ & 9.2 & 88 & bll & MG3 J031849+2135 & N & 1.0 \\
J0322.0+2335 & $0.89_{-0.22}^{+0.24}$ & 20.1 & 122 & bll & MG3 J032201+2336 & N & 5.0 \\
J0331.9+6307 & $0.87_{-0.46}^{+0.45}$ & 3.6 & 65 & bll & GB6 J0331+6307 & N & 1.2 \\
J0333.9+6537 & $0.79_{-0.21}^{+0.25}$ & 16.8 & 96 & bll & TXS 0329+654 & N & 4.9 \\
J0334.2$-$3725 & $1.03_{-0.27}^{+0.28}$ & 13.6 & 77 & bll & PMN J0334-3725 & N & 3.6 \\
J0335.1$-$4459 & $0.99_{-0.51}^{+0.57}$ & 4.3 & 42 & bll & 1RXS J033514.5-445929 & N & 0.6 \\

\hline
\end{tabular}
\end{center}
\end{table*}
\begin{table*}
\begin{center}
\begin{tabular}{lccccrcc}
\hline
4FGL Source Name   &   Redshift Upper Limit & TS$_{peak}$ & HEP (GeV) & Type & Association & TeV & CTA ($\sigma$)\\ 
(1) & (2) & (3)  & (4) & (5) & (6) & (7) & (8) \\ 
\hline
J0338.5+1302 & $0.63_{-0.33}^{+0.27}$ & 2.6 & 66 & bll & RX J0338.4+1302 & N & 4.4 \\
J0338.9$-$2848 & $0.43_{-0.23}^{+0.29}$ & 3.8 & 200 & bcu & NVSS J033859-284619 & N & 3.6 \\
J0343.2$-$6444 & $1.21_{-0.56}^{+0.66}$ & 5.3 & 75 & bll & PMN J0343-6442 & N & 0.2 \\
J0344.4+3432 & $0.41_{-0.23}^{+0.31}$ & 4.0 & 122 & bcu & 1RXS J034424.5+343016 & N & 1.3 \\
J0348.2+6035 & $1.74_{-0.74}^{+0.79}$ & 15.2 & 212 & bcu & NVSS J034819+603509 & N & 0.6 \\
J0350.0+0640 & $1.61_{-0.47}^{+0.68}$ & 35.7 & 112 & bcu & RX J0350.0+0640 & N & 0.0 \\
J0355.3+3909 & $0.39_{-0.28}^{+0.42}$ & 2.2 & 25 & bcu & CRATES J035515+390907 & N & 1.0 \\
J0410.9+4216 & $1.76_{-0.88}^{+1.20}$ & 3.4 & 29 & bcu & B3 0407+421 & N & 0.2 \\
J0420.3$-$6016 & $0.49_{-0.47}^{+0.51}$ & 1.1 & 49 & bcu & 1RXS J042012.8-601446 & N & 0.3 \\
J0423.9+4150 & $2.16_{-0.27}^{+0.28}$ & 199.8 & 49 & bll & 4C +41.11 & N & 6.0 \\
J0425.3+6319 & $0.51_{-0.27}^{+0.27}$ & 3.9 & 115 & bcu & 1RXS J042523.0+632016 & N & 1.9 \\
J0426.7+6826 & $1.84_{-0.80}^{+1.06}$ & 6.4 & 53 & bcu & 4C +68.05 & N & 0.3 \\
J0431.8+7403 & $1.98_{-0.54}^{+0.60}$ & 17.3 & 77 & bll & GB6 J0431+7403 & N & 0.6 \\
J0434.7+0922 & $0.53_{-0.39}^{+0.43}$ & 2.0 & 119 & bll & TXS 0431+092 & N & 1.6 \\
J0439.4$-$3202 & $1.03_{-0.67}^{+0.71}$ & 2.2 & 60 & bcu & 1RXS J043931.4-320045 & N & 0.2 \\
J0442.7+6142 & $0.47_{-0.27}^{+0.29}$ & 3.2 & 92 & bcu & GB6 J0442+6140 & N & 1.7 \\
J0447.2$-$2539 & $0.83_{-0.45}^{+0.52}$ & 4.5 & 78 & bcu & 2MASS J04472149-2539302 & N & 0.7 \\
J0451.8+5721 & $1.25_{-0.82}^{+0.76}$ & 2.2 & 22 & bcu & NVSS J045148+572139 & N & 0.2 \\
J0500.2+5237 & $0.21_{-0.18}^{+0.17}$ & 1.4 & 228 & bcu & GB6 J0500+5238 & N & 3.0 \\
J0501.0+2424 & $1.27_{-0.67}^{+0.85}$ & 3.8 & 43 & bcu & 1RXS J050107.1+242318 & N & 0.3 \\
J0501.7+3048 & $0.49_{-0.29}^{+0.39}$ & 3.4 & 99 & bcu & GB6 J0501+3048 & N & 1.5 \\
J0503.6+4518 & $0.31_{-0.29}^{+0.45}$ & 1.1 & 51 & bcu & GB6 J0503+4517 & N & 1.9 \\
J0506.0+6113 & $0.91_{-0.31}^{+0.38}$ & 17.6 & 141 & bll & RX J0505.9+6113 & N & 2.7 \\
J0515.8+1527 & $1.64_{-0.34}^{+0.39}$ & 35.5 & 67 & bll & GB6 J0515+1527 & N & 3.0 \\
J0525.6$-$6013 & $1.31_{-0.30}^{+0.34}$ & 28.7 & 85 & bcu & SUMSS J052542-601341 & N & 0.8 \\
J0526.7$-$1519 & $0.95_{-0.47}^{+0.66}$ & 5.1 & 103 & bcu & NVSS J052645-151900 & N & 0.9 \\
J0532.0$-$4827 & $0.93_{-0.67}^{+0.56}$ & 1.5 & 49 & bll & PMN J0531-4827 & N & 0.5 \\
J0540.5+5823 & $0.97_{-0.25}^{+0.33}$ & 38.3 & 176 & bll & GB6 J0540+5823 & N & 3.9 \\
J0557.3$-$0615 & $0.61_{-0.56}^{+0.69}$ & 1.2 & 45 & bcu & 1RXS J055717.0-061705 & N & 0.5 \\
J0558.8$-$7459 & $1.74_{-0.70}^{+0.87}$ & 6.3 & 59 & bll & PKS 0600-749 & N & 0.0 \\
J0600.3+1244 & $0.61_{-0.16}^{+0.21}$ & 28.9 & 237 & bcu & NVSS J060015+124344 & N & 4.8 \\
J0601.0+3838 & $0.55_{-0.26}^{+0.36}$ & 6.0 & 112 & bll & B2 0557+38 & N & 3.1 \\
J0601.3+5444 & $0.35_{-0.18}^{+2.12}$ & 4.8 & 28 & bcu & GB6 J0601+5443 & N & 0.3 \\
J0607.4+4739 & $1.96_{-0.34}^{+0.34}$ & 76.6 & 98 & bll & TXS 0603+476 & N & 3.4 \\
J0611.1+4325 & $1.53_{-0.41}^{+0.50}$ & 21.5 & 40 & bcu & 7C 0607+4324 & N & 1.3 \\
J0611.6$-$2712 & $1.19_{-1.18}^{+1.63}$ & 1.0 & -- & bcu & PMN J0611-2709 & N & 0.1 \\
J0612.8+4122 & $1.51_{-0.25}^{+0.29}$ & 30.6 & 132 & bll & B3 0609+413 & N & 5.5 \\
J0617.2+5701 & $2.14_{-0.42}^{+0.53}$ & 43.3 & 72 & bll & 87GB 061258.1+570222 & N & 1.6 \\
J0623.2+3044 & $1.78_{-0.73}^{+1.21}$ & 10.4 & 101 & bll & GB6 J0623+3045 & N & 0.5 \\
J0625.3+4439 & $1.44_{-0.51}^{+0.60}$ & 8.4 & 106 & bll & GB6 J0625+4440 & N & 0.9 \\
J0629.6+2435 & $1.33_{-1.07}^{+1.26}$ & 1.5 & 51 & bcu & GB6 J0629+2437 & N & 0.2 \\
J0647.0$-$5138 & $0.35_{-0.21}^{+0.27}$ & 3.1 & 177 & bcu & 1ES 0646-515 & N & 2.4 \\
J0700.2+1304 & $1.61_{-0.66}^{+0.97}$ & 9.4 & 36 & bll & GB6 J0700+1304 & N & 0.6 \\
J0700.5$-$6610 & $1.53_{-0.27}^{+0.33}$ & 40.8 & 146 & bll & PKS 0700-661 & N & 0.9 \\
J0702.7$-$1951 & $1.96_{-0.49}^{+0.63}$ & 18.0 & 62 & bll & TXS 0700-197 & N & 0.6 \\
J0703.2+6809 & $0.89_{-0.82}^{+1.03}$ & 1.2 & 62 & bcu & GB6 J0703+6808 & N & 0.1 \\
J0705.9+5309 & $0.77_{-0.63}^{+0.77}$ & 1.6 & 13 & bcu & GB6 J0706+5309 & N & 0.3 \\
J0706.1+0246 & $0.49_{-0.17}^{+0.26}$ & 13.1 & 470 & bcu & 1RXS J070609.7+024502 & N & 1.8 \\
J0709.2$-$1527 & $0.81_{-0.26}^{+0.28}$ & 16.6 & 127 & bcu & PKS 0706-15 & N & 4.1 \\
J0718.6$-$4319 & $0.77_{-0.71}^{+0.51}$ & 1.1 & 75 & bll & PMN J0718-4319 & N & 1.4 \\
J0721.3$-$0222 & $1.07_{-0.59}^{+0.70}$ & 2.8 & 42 & bll & 1RXS J072114.5-022047 & N & 0.4 \\
J0723.0$-$0732 & $0.61_{-0.34}^{+0.37}$ & 3.6 & 147 & bll & 1RXS J072259.5-073131 & N & 1.9 \\
J0725.5+0216 & $0.99_{-0.65}^{+0.69}$ & 2.1 & 44 & bcu & NVSS J072534+021645 & N & 0.5 \\
J0733.5$-$5445 & $1.72_{-0.96}^{+1.26}$ & 2.9 & 15 & bcu & SUMSS J073334-544544 & N & 0.0 \\
J0737.3$-$8247 & $1.55_{-0.44}^{+0.51}$ & 17.6 & 80 & bcu & SUMSS J073706-824836 & N & 0.0 \\
J0746.3$-$0225 & $0.79_{-0.40}^{+0.56}$ & 6.1 & 85 & bcu & 2MASS J07462703-0225492 & N & 1.6 \\
J0746.6$-$4754 & $0.89_{-0.20}^{+0.23}$ & 25.1 & 103 & bll & PMN J0746-4755 & N & 6.4 \\
J0747.3$-$3310 & $1.38_{-0.71}^{+0.88}$ & 3.6 & 44 & bll & PKS 0745-330 & N & 0.3 \\
J0747.5$-$4927 & $0.37_{-0.27}^{+0.31}$ & 2.0 & 19 & bcu & 2MASS J07472476-4926332 & N & 2.5 \\
J0749.0$-$2956 & $1.01_{-0.90}^{+0.99}$ & 1.3 & 51 & bcu & NVSS J074913-295658 & N & 0.1 \\
J0754.0+0451 & $1.27_{-0.55}^{+0.67}$ & 5.3 & 24 & bcu & GB6 J0754+0452 & N & 0.6 \\

\hline
\end{tabular}
\end{center}
\end{table*}
\begin{table*}
\begin{center}
\begin{tabular}{lccccrcc}
\hline
4FGL Source Name   &   Redshift Upper Limit & TS$_{peak}$ & HEP (GeV) & Type & Association & TeV & CTA ($\sigma$)\\ 
(1) & (2) & (3)  & (4) & (5) & (6) & (7) & (8) \\ 
\hline
J0756.3$-$6431 & $0.67_{-0.51}^{+0.58}$ & 1.8 & 40 & bll & SUMSS J075625-643031 & N & 0.2 \\
J0802.3$-$0942 & $0.89_{-0.39}^{+0.48}$ & 6.4 & 68 & bcu & NVSS J080216-094215 & N & 1.4 \\
J0806.9$-$2151 & $2.68_{-1.64}^{+0.30}$ & 5.4 & 24 & bcu & TXS 0804-217 & N & 0.1 \\
J0811.0$-$7529 & $0.49_{-0.30}^{+0.25}$ & 1.9 & 94 & bll & PMN J0810-7530 & N & 0.0 \\
J0812.0+0237 & $0.21_{-0.15}^{+0.18}$ & 2.0 & 149 & bll & PMN J0811+0237 & N & 4.3 \\
J0814.2$-$1013 & $1.48_{-0.57}^{+0.68}$ & 8.5 & 79 & bll & NVSS J081411-101208 & N & 0.7 \\
J0826.4$-$6404 & $0.17_{-0.16}^{+0.20}$ & 1.0 & 77 & bll & SUMSS J082627-640414 & N & 4.3 \\
J0829.6$-$1140 & $1.86_{-0.56}^{+0.81}$ & 15.6 & 29 & bcu & NVSS J082939-114103 & N & 0.4 \\
J0829.7+5105 & $1.03_{-0.51}^{+0.64}$ & 5.9 & 35 & bll & GB6 J0829+5108 & N & 0.5 \\
J0849.1+6607 & $1.82_{-0.53}^{+0.72}$ & 19.7 & -- & bll & GB6 J0848+6605 & N & 0.8 \\
J0849.4$-$2911 & $1.44_{-0.82}^{+1.04}$ & 3.0 & 45 & bcu & NVSS J084922-291149 & N & 0.1 \\
J0853.1$-$3657 & $0.93_{-0.29}^{+0.29}$ & 12.1 & 106 & bcu & NVSS J085310-365820 & N & 3.2 \\
J0856.4$-$5309 & $1.38_{-0.91}^{+1.30}$ & 2.7 & 74 & bcu & PMN J0856-5312 & N & 0.1 \\
J0905.6+1358 & $1.03_{-0.33}^{+0.34}$ & 11.5 & 125 & bll & MG1 J090534+1358 & N & 3.8 \\
J0929.3+5014 & $0.45_{-0.39}^{+0.47}$ & 1.4 & -- & bll & GB6 J0929+5013 & N & 1.4 \\
J0935.3$-$1736 & $1.44_{-0.62}^{+0.98}$ & 9.1 & -- & bcu & NVSS J093514-173658 & N & 0.4 \\
J0936.3$-$2111 & $0.99_{-0.95}^{+0.71}$ & 1.0 & -- & bcu & TXS 0933-209 & N & 0.2 \\
J0952.6$-$5048 & $2.30_{-1.64}^{+0.68}$ & 3.2 & -- & bcu & 2MASS J09524301-5049538 & N & 0.1 \\
J0953.4$-$7659 & $0.71_{-0.31}^{+0.37}$ & 8.1 & 101 & bcu & RX J0953.1-7657 & N & 0.0 \\
J1018.1+1905 & $1.15_{-0.72}^{+0.85}$ & 2.1 & 34 & bll & NVSS J101808+190614 & N & 0.3 \\
J1022.4$-$4231 & $1.53_{-1.18}^{+1.43}$ & 1.5 & -- & bll & PMN J1022-4232 & N & 0.0 \\
J1027.0$-$8542 & $2.10_{-0.27}^{+0.29}$ & 362.2 & 105 & bll & PKS 1029-85 & N & 0.0 \\
J1027.6+8251 & $1.50_{-0.94}^{+1.41}$ & 2.8 & 50 & bcu & 2MASS J10284195+8253398 & N & 0.0 \\
J1041.1$-$1201 & $0.59_{-0.42}^{+0.77}$ & 2.3 & 59 & bcu & NVSS J104108-120332 & N & 0.7 \\
J1055.5$-$0125 & $0.47_{-0.39}^{+0.43}$ & 1.5 & 107 & bll & RX J1055.5-0126 & N & 1.4 \\
J1107.6+0222 & $2.06_{-0.57}^{+0.74}$ & 17.3 & 35 & bll & NVSS J110735+022225 & N & 0.7 \\
J1110.2+7135 & $0.43_{-0.28}^{+0.32}$ & 2.5 & 65 & bll & RX J1110.5+7133 & N & 1.5 \\
J1123.8+7230 & $0.53_{-0.25}^{+0.38}$ & 6.3 & 61 & bll & RX J1123.8+7230 & N & 1.1 \\
J1124.6$-$0809 & $0.67_{-0.66}^{+1.00}$ & 1.1 & 29 & bcu & AT20G J112437-080643 & N & 0.2 \\
J1125.1$-$2101 & $1.76_{-0.62}^{+0.90}$ & 12.6 & 50 & bll & PMN J1125-2100 & N & 0.4 \\
J1141.5$-$1408 & $0.77_{-0.54}^{+0.71}$ & 2.3 & 23 & bll & 1RXS J114142.2-140757 & N & 0.5 \\
J1155.5$-$3418 & $0.81_{-0.37}^{+0.48}$ & 7.0 & 126 & bcu & NVSS J115520-341718 & N & 1.4 \\
J1156.6$-$2248 & $1.44_{-0.78}^{+0.99}$ & 3.1 & 31 & bcu & NVSS J115633-225004 & N & 0.1 \\
J1220.1+3432 & $1.35_{-0.80}^{+1.11}$ & 3.1 & 43 & bll & GB2 1217+348 & N & 0.3 \\
J1223.5+0818 & $0.87_{-0.60}^{+0.72}$ & 2.2 & 39 & bcu & SDSS J122327.49+082030.4 & N & 0.4 \\
J1232.5$-$3720 & $0.49_{-0.34}^{+0.38}$ & 2.2 & 52 & bcu & NVSS J123235-372051 & N & 1.7 \\
J1233.7$-$0144 & $0.57_{-0.37}^{+0.39}$ & 2.3 & 79 & bll & NVSS J123341-014426 & N & 1.7 \\
J1240.1$-$6846 & $1.50_{-1.39}^{+1.50}$ & 1.1 & -- & bcu & 2MASS J12400694-6844532 & N & 0.0 \\
J1240.4$-$7148 & $0.65_{-0.22}^{+0.33}$ & 15.0 & 348 & bcu & 2MASS J12404205-7147599 & N & 2.9 \\
J1248.3+5820 & $2.68_{-0.18}^{+0.18}$ & 1672.9 & 198 & bll & PG 1246+586 & N & 12.7 \\
J1249.2$-$2809 & $1.13_{-0.52}^{+0.57}$ & 5.4 & 29 & bcu & GALEXASC J124926.84-280857.5 & N & 0.6 \\
J1253.2+5301 & $0.81_{-0.32}^{+0.32}$ & 5.8 & 155 & bll & S4 1250+53 & N & 4.1 \\
J1256.1$-$5919 & $0.49_{-0.45}^{+0.49}$ & 1.2 & 52 & bcu & PMN J1256-5919 & N & 0.5 \\
J1259.8$-$3749 & $0.75_{-0.45}^{+0.51}$ & 3.0 & 90 & bll & NVSS J125949-374856 & N & 1.0 \\
J1304.2$-$2412 & $1.96_{-0.57}^{+0.79}$ & 17.1 & 34 & bll & PMN J1304-2412 & N & 0.4 \\
J1307.6$-$4259 & $0.57_{-0.17}^{+0.20}$ & 13.4 & 114 & bll & 1RXS J130737.8-425940 & N & 9.2 \\
J1311.8+3954 & $1.25_{-1.01}^{+1.19}$ & 1.6 & -- & bll & FIRST J131146.0+395317 & N & 0.1 \\
J1327.8+2522 & $0.73_{-0.66}^{+0.98}$ & 1.3 & 12 & bll & NVSS J132758+252750 & N & 0.2 \\
J1328.5$-$4727 & $0.77_{-0.35}^{+0.36}$ & 5.0 & 99 & bll & 1WGA J1328.6-4727 & N & 1.7 \\
J1330.2+7002 & $1.46_{-0.48}^{+0.57}$ & 11.0 & 49 & bll & NVSS J133025+700141 & N & 0.6 \\
J1338.9+1153 & $0.75_{-0.43}^{+0.84}$ & 4.2 & 23 & bll & SDSS J133859.05+115316.7 & N & 1.0 \\
J1347.1$-$2959 & $1.59_{-0.63}^{+0.85}$ & 9.4 & 56 & bll & NVSS J134706-295840 & N & 0.3 \\
J1349.5$-$7153 & $0.67_{-0.55}^{+0.82}$ & 1.6 & -- & bcu & 2MASS J13490974-7152510 & N & 0.1 \\
J1351.3+1115 & $1.31_{-0.38}^{+0.42}$ & 15.8 & 41 & bll & RX J1351.3+1115 & N & 1.9 \\
J1353.6$-$6640 & $0.31_{-0.17}^{+0.16}$ & 3.5 & 127 & bll & 1RXS J135341.1-664002 & N & 5.6 \\
J1406.1$-$2508 & $0.87_{-0.63}^{+0.78}$ & 2.0 & 68 & bll & NVSS J140609-250808 & N & 0.4 \\
J1407.6$-$4301 & $1.64_{-0.92}^{+1.27}$ & 2.9 & 18 & bcu & SUMSS J140739-430231 & N & 0.1 \\
J1411.5$-$0723 & $0.33_{-0.27}^{+1.05}$ & 1.7 & 99 & bcu & NVSS J141133-072252 & N & 0.7 \\
J1420.3+0612 & $0.93_{-0.69}^{+1.41}$ & 2.5 & -- & bll & SDSS J142013.69+061428.6 & N & 0.2 \\
J1424.6+1447 & $1.35_{-0.66}^{+0.94}$ & 4.9 & -- & bll & SDSS J142436.29+144910.5 & N & 0.4 \\
J1427.7$-$3215 & $0.85_{-0.46}^{+0.54}$ & 3.9 & 98 & bll & NVSS J142750-321515 & N & 0.9 \\

\hline
\end{tabular}
\end{center}
\end{table*}
\begin{table*}
\begin{center}
\begin{tabular}{lccccrcc}
\hline
4FGL Source Name   &   Redshift Upper Limit & TS$_{peak}$ & HEP (GeV) & Type & Association & TeV & CTA ($\sigma$)\\ 
(1) & (2) & (3)  & (4) & (5) & (6) & (7) & (8) \\ 
\hline
J1432.2+5051 & $1.07_{-0.95}^{+1.71}$ & 1.4 & -- & bcu & NVSS J143217+505603 & N & 0.1 \\
J1434.8+6640 & $0.45_{-0.41}^{+0.57}$ & 1.3 & 151 & bll & 1RXS J143442.0+664031 & N & 0.5 \\
J1440.0$-$2343 & $0.81_{-0.41}^{+0.48}$ & 4.4 & 22 & bcu & PMN J1439-2341 & N & 1.1 \\
J1446.8$-$1830 & $1.51_{-0.91}^{+1.23}$ & 3.2 & 25 & bcu & NVSS J144644-182922 & N & 0.1 \\
J1448.0+3608 & $1.76_{-0.20}^{+0.23}$ & 263.3 & 140 & bll & RBS 1432 & N & 6.9 \\
J1454.4+5124 & $1.29_{-0.25}^{+0.26}$ & 22.4 & 78 & bll & TXS 1452+516 & N & 4.3 \\
J1455.8$-$7601 & $1.31_{-0.75}^{+0.91}$ & 4.1 & 81 & bcu & SUMSS J145543-760054 & N & 0.0 \\
J1457.8$-$4642 & $0.71_{-0.39}^{+0.91}$ & 5.2 & 55 & bcu & PMN J1457-4642 & N & 0.6 \\
J1511.8$-$0513 & $1.23_{-0.34}^{+0.37}$ & 18.6 & 56 & bll & NVSS J151148-051345 & N & 2.0 \\
J1516.9+1934 & $2.88_{-2.16}^{+0.10}$ & 1.7 & -- & bll & PKS 1514+197 & N & 0.1 \\
J1518.0$-$2731 & $0.89_{-0.31}^{+0.31}$ & 13.0 & 65 & bll & TXS 1515-273 & Y & 4.2 \\
J1520.0$-$0905 & $1.44_{-0.65}^{+0.89}$ & 5.7 & 41 & bcu & 1RXS J151959.7-090434 & N & 0.2 \\
J1528.2$-$2905 & $0.53_{-0.49}^{+1.32}$ & 1.2 & -- & bcu & 2MASS J15282165-2858132 & N & 0.2 \\
J1532.7$-$1319 & $1.72_{-0.56}^{+0.62}$ & 6.2 & 46 & fsrq & TXS 1530-131 & N & 0.8 \\
J1535.3$-$3135 & $1.82_{-0.72}^{+1.15}$ & 11.0 & -- & bcu & 2MASS J15352963-3133461 & N & 0.2 \\
J1537.7$-$7957 & $1.70_{-0.70}^{+0.88}$ & 5.2 & 34 & bcu & PMN J1537-7958 & N & 0.0 \\
J1537.9$-$1344 & $1.17_{-0.46}^{+0.58}$ & 13.7 & 36 & bcu & 1RXS J153757.1-134334 & N & 0.8 \\
J1539.7$-$1127 & $0.57_{-0.41}^{+0.45}$ & 2.0 & 116 & bll & PMN J1539-1128 & N & 1.6 \\
J1549.8$-$0659 & $0.87_{-0.30}^{+0.32}$ & 10.7 & 127 & bcu & NVSS J154952-065907 & N & 2.9 \\
J1553.5$-$3118 & $1.03_{-0.28}^{+0.29}$ & 21.8 & 123 & bll & 1RXS J155333.4-311841 & N & 4.2 \\
J1602.9$-$1928 & $1.17_{-0.76}^{+1.35}$ & 3.1 & 15 & bcu & 2MASS J16024856-1929470 & N & 0.2 \\
J1603.8$-$4903 & $0.17_{-0.15}^{+0.13}$ & 1.2 & 133 & bll & PMN J1603-4904 & N & 21.1 \\
J1610.7$-$6648 & $1.46_{-0.15}^{+0.14}$ & 125.8 & 124 & bll & PMN J1610-6649 & N & 4.3 \\
J1624.6+5651 & $1.27_{-1.04}^{+1.15}$ & 1.4 & 24 & bll & SBS 1623+569 & N & 0.1 \\
J1632.2+0854 & $1.50_{-1.01}^{+1.48}$ & 2.0 & 33 & bcu & NVSS J163211+085608 & N & 0.1 \\
J1637.7+7326 & $1.51_{-0.52}^{+0.84}$ & 18.3 & 75 & bll & RX J1637.9+7326 & N & 0.4 \\
J1637.8$-$3449 & $1.33_{-0.31}^{+0.35}$ & 27.4 & 64 & bll & NVSS J163750-344915 & N & 2.6 \\
J1646.7$-$1330 & $0.61_{-0.50}^{+0.57}$ & 1.5 & 64 & bcu & 2MASS J16465897-1329454 & N & 0.6 \\
J1647.4$-$6438 & $1.09_{-0.64}^{+0.83}$ & 3.0 & 66 & bcu & PMN J1647-6437 & N & 0.1 \\
J1649.4+5235 & $1.94_{-0.59}^{+0.68}$ & 7.7 & 61 & bll & 87GB 164812.2+524023 & N & 0.7 \\
J1650.3$-$5045 & $1.96_{-0.57}^{+0.67}$ & 13.0 & 67 & bcu & PMN J1650-5044 & N & 0.5 \\
J1651.6+7219 & $0.47_{-0.31}^{+0.32}$ & 2.6 & 134 & bll & RX J1651.6+7218 & N & 1.1 \\
J1659.7$-$3131 & $0.85_{-0.58}^{+0.76}$ & 2.8 & 18 & bcu & NVSS J165949-313047 & N & 0.9 \\
J1704.5$-$0527 & $0.41_{-0.23}^{+0.25}$ & 3.4 & 165 & bll & NVSS J170433-052839 & N & 4.5 \\
J1735.4$-$1118 & $1.13_{-0.82}^{+0.74}$ & 1.6 & 30 & bcu & PMN J1735-1117 & N & 0.2 \\
J1735.8$-$5932 & $0.83_{-0.62}^{+0.82}$ & 1.9 & 60 & bcu & 1RXS J173553.1-593205 & N & 0.1 \\
J1741.2+5739 & $1.13_{-0.95}^{+1.23}$ & 1.4 & -- & bcu & NVSS J174111+573812 & N & 0.1 \\
J1744.4+1851 & $0.91_{-0.41}^{+0.64}$ & 11.6 & 100 & bcu & 1RXS J174420.1+185215 & N & 0.8 \\
J1744.6$-$5713 & $1.78_{-0.57}^{+0.79}$ & 15.1 & 41 & bll & PMN J1744-5715 & N & 0.1 \\
J1744.9$-$1727 & $0.21_{-0.16}^{+0.17}$ & 1.8 & 197 & bcu & 1RXS J174459.5-172640 & N & 8.5 \\
J1748.1+2702 & $1.25_{-1.08}^{+1.07}$ & 1.3 & 31 & bcu & 87GB 174618.6+270457 & N & 0.1 \\
J1759.1$-$4822 & $1.40_{-1.33}^{+1.24}$ & 1.0 & 15 & bcu & PMN J1758-4820 & N & 0.0 \\
J1800.1+2812 & $1.40_{-1.06}^{+1.04}$ & 1.6 & 37 & bcu & NVSS J180002+281050 & N & 0.1 \\
J1800.1+7037 & $1.53_{-0.68}^{+0.95}$ & 6.4 & 56 & bll & RX J1759.8+7037 & N & 0.3 \\
J1809.7+2910 & $1.42_{-0.44}^{+0.48}$ & 10.6 & 62 & bll & MG2 J180948+2910 & N & 1.6 \\
J1810.7+5335 & $2.02_{-0.64}^{+0.88}$ & 12.7 & 19 & bcu & 2MASS J18103800+5335016 & N & 0.3 \\
J1811.0+1608 & $0.63_{-0.62}^{+1.16}$ & 1.0 & 10 & bll & 87GB 180835.5+160714 & N & 0.4 \\
J1820.3+3624 & $1.40_{-1.15}^{+0.84}$ & 3.4 & 55 & bll & NVSS J182021+362343 & N & 0.4 \\
J1823.6$-$3453 & $0.29_{-0.12}^{+0.12}$ & 7.0 & 177 & bcu & NVSS J182338-345412 & N & 23.9 \\
J1829.3+5402 & $1.38_{-0.38}^{+0.54}$ & 20.2 & 66 & bll & RX J1829.3+5403 & N & 1.2 \\
J1830.0+1324 & $1.50_{-0.74}^{+0.90}$ & 4.0 & 55 & bll & MG1 J183001+1323 & N & 0.4 \\
J1832.6$-$5658 & $1.05_{-0.71}^{+1.39}$ & 3.3 & 41 & bll & PMN J1832-5659 & N & 0.1 \\
J1841.3+2909 & $0.27_{-0.23}^{+0.28}$ & 1.4 & 58 & bll & MG3 J184126+2910 & N & 2.2 \\
J1844.4+1547 & $0.93_{-0.34}^{+0.35}$ & 7.4 & 162 & bll & NVSS J184425+154646 & N & 3.8 \\
J1845.0+1613 & $0.99_{-0.87}^{+0.85}$ & 1.3 & 16 & bcu & 87GB 184225.9+161105 & N & 0.3 \\
J1846.7+7238 & $1.13_{-0.58}^{+0.71}$ & 4.1 & 42 & bcu & RX J1846.1+7237 & N & 0.3 \\
J1850.5+2631 & $0.83_{-0.26}^{+0.32}$ & 20.7 & 138 & bll & NVSS J185023+263151 & N & 1.9 \\
J1858.5+0640 & $1.25_{-0.86}^{+0.77}$ & 2.8 & 54 & bcu & NVSS J185831+064016 & N & 0.8 \\
J1859.0+2329 & $0.63_{-0.62}^{+0.81}$ & 1.0 & -- & bcu & NVSS J185857+233007 & N & 0.4 \\
J1903.2+5540 & $0.83_{-0.22}^{+0.21}$ & 11.2 & 144 & bll & TXS 1902+556 & N & 7.1 \\
J1909.5+3511 & $2.92_{-1.25}^{+0.04}$ & 15.9 & 57 & bcu & TXS 1907+350 & N & 0.2 \\

\hline
\end{tabular}
\end{center}
\end{table*}
\begin{table*}
\begin{center}
\begin{tabular}{lccccrcc}
\hline
4FGL Source Name   &   Redshift Upper Limit & TS$_{peak}$ & HEP (GeV) & Type & Association & TeV & CTA ($\sigma$)\\ 
(1) & (2) & (3)  & (4) & (5) & (6) & (7) & (8) \\ 
\hline
J1912.1$-$0803 & $0.87_{-0.72}^{+0.74}$ & 1.4 & 75 & bcu & PMN J1912-0804 & N & 0.4 \\
J1913.9+4439 & $1.94_{-0.66}^{+0.87}$ & 5.6 & 34 & bll & 1RXS J191401.9+443849 & N & 0.4 \\
J1918.1+3752 & $0.91_{-0.73}^{+0.81}$ & 1.6 & 54 & bcu & 1RXS J191810.2+375315 & N & 0.3 \\
J1921.3$-$1231 & $1.72_{-0.54}^{+0.65}$ & 11.1 & 63 & bll & TXS 1918-126 & N & 0.5 \\
J1925.0+2815 & $0.39_{-0.28}^{+0.38}$ & 2.2 & 46 & bcu & NVSS J192502+281542 & N & 2.0 \\
J1925.8$-$2220 & $0.43_{-0.40}^{+0.46}$ & 1.2 & 104 & bll & TXS 1922-224 & N & 1.0 \\
J1926.8+6154 & $1.17_{-0.19}^{+0.18}$ & 81.8 & 78 & bll & 87GB 192614.4+614823 & N & 6.7 \\
J1927.5+6117 & $1.29_{-0.40}^{+0.48}$ & 12.5 & 61 & bll & S4 1926+61 & N & 1.2 \\
J1933.3+0726 & $0.61_{-0.18}^{+0.21}$ & 20.8 & 189 & bll & 1RXS J193320.3+072616 & N & 7.7 \\
J1942.5$-$5827 & $0.77_{-0.75}^{+0.86}$ & 1.1 & -- & bcu & SUMSS J194224-582824 & N & 0.1 \\
J1942.7+1033 & $0.55_{-0.12}^{+0.12}$ & 35.6 & 180 & bll & 87GB 194024.3+102612 & N & 18.5 \\
J1944.4$-$4523 & $0.61_{-0.27}^{+0.32}$ & 6.5 & 50 & bcu & 1RXS J194422.6-452326 & N & 2.4 \\
J1946.0+0937 & $1.46_{-0.55}^{+0.72}$ & 12.3 & 97 & bcu & 87GB 194328.9+092856 & N & 0.7 \\
J1949.5+0906 & $1.31_{-0.43}^{+0.52}$ & 12.9 & 50 & bll & 1RXS J194934.1+090655 & N & 1.3 \\
J1955.1$-$1604 & $1.13_{-0.38}^{+0.43}$ & 13.3 & 40 & bll & 1RXS J195500.6-160328 & N & 1.6 \\
J2001.2+4353 & $2.48_{-0.33}^{+0.30}$ & 336.4 & 155 & bll & MG4 J200112+4352 & Y & 5.5 \\
J2002.4$-$7119 & $0.53_{-0.27}^{+0.32}$ & 4.6 & 113 & bcu & SUMSS J200227-711940 & N & 1.3 \\
J2002.6+6302 & $1.90_{-0.52}^{+0.68}$ & 30.5 & 98 & bcu & 1RXS J200245.4+630226 & N & 0.5 \\
J2005.1+7003 & $0.95_{-0.45}^{+0.42}$ & 2.9 & 77 & bll & 1RXS J200504.0+700445 & N & 1.6 \\
J2012.0+4629 & $0.81_{-0.21}^{+0.21}$ & 12.5 & 89 & bll & 7C 2010+4619 & N & 8.6 \\
J2024.4$-$0847 & $0.51_{-0.47}^{+0.49}$ & 1.2 & 28 & bll & 1RXS J202428.9-084810 & N & 1.1 \\
J2025.3$-$2231 & $0.55_{-0.42}^{+0.57}$ & 1.8 & 62 & bcu & NVSS J202515-223016 & N & 1.0 \\
J2026.1+7645 & $0.51_{-0.20}^{+0.23}$ & 11.0 & 192 & bcu & 1RXS J202633.4+764432 & N & 2.7 \\
J2026.6+3449 & $1.57_{-0.62}^{+0.78}$ & 9.2 & 28 & bcu & NVSS J202638+345022 & N & 0.8 \\
J2035.9+4901 & $0.63_{-0.43}^{+0.45}$ & 2.2 & 99 & bcu & 2MASS J20355146+4901490 & N & 1.1 \\
J2039.0$-$1046 & $0.99_{-0.89}^{+0.62}$ & 1.1 & 38 & bll & TXS 2036-109 & N & 0.3 \\
J2040.1$-$4621 & $1.13_{-0.59}^{+0.81}$ & 5.3 & 37 & bcu & 2MASS J20400660-4620180 & N & 0.4 \\
J2041.8$-$7319 & $1.57_{-0.51}^{+0.64}$ & 13.4 & 52 & bcu & SUMSS J204201-731911 & N & 0.1 \\
J2045.1$-$2346 & $1.09_{-0.50}^{+0.71}$ & 7.4 & 13 & bcu & NVSS J204457-234643 & N & 0.6 \\
J2047.9$-$3122 & $1.03_{-0.77}^{+1.02}$ & 1.8 & -- & bcu & NVSS J204806-312016 & N & 0.2 \\
J2056.7+4939 & $0.17_{-0.07}^{+0.06}$ & 6.4 & 309 & bcu & RGB J2056+496 & Y & 18.7 \\
J2102.3+4702 & $1.42_{-0.57}^{+0.64}$ & 5.4 & 54 & bcu & MG4 J210218+4702 & N & 0.7 \\
J2104.3$-$0212 & $0.75_{-0.26}^{+0.36}$ & 19.5 & 113 & bll & NVSS J210421-021239 & N & 3.3 \\
J2108.9$-$6638 & $1.78_{-0.58}^{+0.76}$ & 8.6 & 57 & bll & PKS 2104-668 & N & 0.1 \\
J2109.8$-$8618 & $0.33_{-0.32}^{+0.35}$ & 1.1 & 88 & bcu & 1RXS J210959.5-861853 & N & 0.0 \\
J2110.3+0404 & $1.27_{-1.22}^{+1.39}$ & 1.1 & 24 & bcu & NVSS J211019+040418 & N & 0.1 \\
J2115.2+1218 & $0.57_{-0.43}^{+0.45}$ & 1.9 & 119 & bcu & NVSS J211522+121802 & N & 0.9 \\
J2126.1$-$3922 & $0.97_{-0.80}^{+1.04}$ & 1.7 & -- & bcu & PMN J2126-3921 & N & 0.2 \\
J2126.5+1842 & $1.05_{-0.77}^{+0.87}$ & 3.0 & -- & bcu & 87GB 212407.5+182753 & N & 0.5 \\
J2127.7+3612 & $0.53_{-0.47}^{+0.43}$ & 1.2 & 37 & bll & B2 2125+35 & N & 1.8 \\
J2133.9+6646 & $2.26_{-0.55}^{+0.71}$ & 30.1 & 65 & bll & NVSS J213349+664706 & N & 0.8 \\
J2139.4$-$4235 & $1.96_{-0.23}^{+0.22}$ & 206.2 & 152 & bll & MH 2136-428 & N & 6.1 \\
J2142.1+4501 & $1.53_{-0.88}^{+1.21}$ & 4.2 & -- & bcu & B3 2140+447 & N & 0.2 \\
J2142.4+3659 & $0.77_{-0.36}^{+0.42}$ & 5.7 & 25 & bcu & 2MASS J21422658+3659481 & N & 1.1 \\
J2144.2+3132 & $0.79_{-0.74}^{+0.78}$ & 1.1 & 85 & bll & MG3 J214415+3132 & N & 0.4 \\
J2156.0+1818 & $0.79_{-0.30}^{+0.34}$ & 12.5 & 215 & bll & RX J2156.0+1818 & N & 2.7 \\
J2210.8+3203 & $0.51_{-0.23}^{+0.35}$ & 7.9 & 347 & bcu & 1RXS J221058.3+320327 & N & 0.8 \\
J2213.5$-$4754 & $1.53_{-0.76}^{+0.95}$ & 4.8 & 28 & bcu & SUMSS J221330-475426 & N & 0.2 \\
J2215.4+0544 & $1.40_{-0.91}^{+1.16}$ & 2.4 & -- & bcu & NVSS J221513+054454 & N & 0.2 \\
J2229.1+2254 & $1.82_{-0.73}^{+1.15}$ & 11.2 & 49 & bcu & NVSS J222913+225511 & N & 0.2 \\
J2247.8+4413 & $0.37_{-0.15}^{+0.17}$ & 7.0 & 169 & bll & RGB J2247+442 & N & 4.6 \\
J2300.3+3136 & $1.68_{-0.74}^{+0.79}$ & 3.2 & 45 & bll & NVSS J230022+313703 & N & 0.8 \\
J2300.8$-$0736 & $0.35_{-0.31}^{+0.40}$ & 1.3 & 157 & bcu & 2MASS J23005469-0735438 & N & 1.3 \\
J2304.6+3704 & $0.93_{-0.25}^{+0.28}$ & 21.5 & 153 & bll & 1RXS J230437.1+370506 & N & 3.3 \\
J2313.4$-$6922 & $1.50_{-0.57}^{+0.81}$ & 14.3 & 17 & bcu & SUMSS J231347-692332 & N & 0.1 \\
J2321.7$-$6438 & $0.71_{-0.47}^{+0.46}$ & 2.2 & 53 & bcu & PMN J2321-6438 & N & 0.3 \\
J2329.7+6101 & $0.65_{-0.27}^{+0.32}$ & 7.1 & 143 & bcu & NVSS J232938+610113 & N & 2.5 \\
J2339.2$-$7403 & $0.43_{-0.40}^{+0.46}$ & 1.2 & 86 & bcu & 1RXS J233919.8-740439 & N & 0.6 \\
J2351.3$-$7559 & $0.77_{-0.45}^{+0.55}$ & 3.7 & 134 & bll & SUMSS J235115-760012 & N & 0.0 \\

\hline
\end{tabular}
\end{center}
\caption{Column information are as follows: (1) 4FGL source name; (2) redshift upper limit, i.e.,~ the redshift at TS$_{peak}$ derived from our EBL-attenuation methodology, uncertainties are at 68\% C.L.; (3) value of the peak of the Test Statistic profile; (4) highest enegy photon; (5) source type (bll is BL Lac and bcu is blazar candidate of uncertain type); (6) source association; (7) whether the source is detected by IACTs, Yes or No; (8) predicted significance of CTA detection in 20h, either from CTA-North (positive declination) or CTA-South (negative declination) depending on the observability of the source.}\label{tab1}

\end{table*}






\bsp	
\label{lastpage}
\end{document}